\documentclass[12pt]{article}
\usepackage{amsmath}
\usepackage{graphicx}
\usepackage{caption}
\usepackage{natbib}
\usepackage{amssymb}
\usepackage{url} 
\usepackage{pdflscape}
\usepackage{color}
\usepackage{verbatim}
\newcommand{\blind}{0}
\usepackage{hyperref}
\hypersetup{
	colorlinks=blue,
	linkcolor=blue,
	filecolor=blue,      
	urlcolor=blue,
	pdftitle={Overleaf Example},
	pdfpagemode=FullScreen,
}

\begin{document}

\def\spacingset#1{\renewcommand{\baselinestretch}%
{#1}\small\normalsize} \spacingset{1}


\if0\blind
{
  \title{\bf Entropy Adjusted Graphical Lasso for Sparse Precision Matrix Estimation }
  \author{Vahe Avagyan\thanks{Biometris, Wageningen University and Research, Wageningen, The Netherlands  }}
  \maketitle
} \fi

\if1\blind
{
  \bigskip
  \bigskip
  \bigskip
  \begin{center}
    {\LARGE\bf Title}
\end{center}
  \medskip
} \fi

\bigskip
\begin{abstract}
The estimation of a precision matrix is a crucial problem in various research fields, particularly when working with high dimensional data. In such settings, the most common approach is to use the penalized maximum likelihood. The literature typically employs Lasso, Ridge and Elastic Net norms, which effectively shrink the entries of the estimated precision matrix. Although these shrinkage approaches provide well-conditioned precision matrix estimates, they do not explicitly address the uncertainty associated with these estimated matrices. In fact, as the matrix becomes sparser, the precision matrix imposes fewer restrictions, leading to greater variability in the distribution, and thus, to higher entropy. In this paper, we introduce an entropy-adjusted extension of widely used Graphical Lasso using an additional log-determinant penalty term. The objective of the proposed technique is to impose sparsity on the precision matrix estimate and adjust the uncertainty through the log-determinant term.  The advantage of the proposed method compared to the existing ones in the literature is evaluated through comprehensive numerical analyses, including both simulated and real-world datasets. The results demonstrate its benefits compared to existing approaches in the literature, with respect to several evaluation metrics.

\end{abstract}

\noindent%
\vfill

\newpage
\spacingset{1.75} 

\section{Introduction}
\label{section:1}

\indent

Inverse of the covariance matrix (also known as precision matrix) has an important role in modern statistics and machine learning. Its accurate estimation is required in different research fields including finance \citep{choi2019high}, psychology \citep{till2023network}, genetics \citep{cai2013covariate}, brain studies \citep{huang}, etc. The estimated precision matrix is fundamental in discriminant analysis, forecasting and several other statistical methodologies \citep{mclachlan}. 

Precision matrix is closely related to the partial correlations among variables. Suppose the true precision matrix is $\Omega=[\omega_{ij}]_{1\le i,j\le p} \in {\mathbb{R}}^{p\times p}$, which is positive definite. The entries of partial correlation matrix $P=[\rho_{ij}]_{1\le i,j\le p}$ can be written in terms of the precision matrix entries:
\begin{eqnarray}
	\label{partialcorr}
	\rho_{ij} &=& - \dfrac{\omega_{ij}}{\sqrt{\omega_{ii} \omega_{jj}}}
\end{eqnarray}

Therefore, the ${(i,j)}$ entry of the precision matrix is zero if and only if the partial correlation of the variables $X^i$ and $X^j$ is zero. Under the assumption of multivariate normality, the entry $\omega_{ij}=0$ of the precision matrix indicates the conditional independence between the variables $X^i$ and $X^j$, given all the other variables \citep{dempster}. In this way, the precision matrix is closely related to the Gaussian Graphical Models (GGM) which can represent the conditional independence of multivariate normally distributed variables in a convenient form as a graph (e.g., gene interaction networks). The GGM is an undirected graph defined as ${G=(N,E)}$. Here, the set of nodes ${N=\{1,...,p\}}$ represents the variables and the set of edges $E\subseteq N \times N$ consists of the pair indexes ${(i,j)}$ corresponding to the `active' entries ${\omega_{ij}\not = 0}$ for $1\le i,j \le p$ \citep{lauritzen}. The selection of GGM is an important problem in studying genetic interactions, human brain activity network, psychological networks, etc.

 In recent decades, substantial research has focused on estimating the precision matrix under high dimensional settings. The most popular approach under high dimensional settings is applying Lasso (Least absolute shrinkage and selection operator) or $\ell_1$ norm penalty on an objective loss function. This approach is proposed in the regression context by \cite{tibshirani} and is widely used for estimating precision and covariance matrices. In particular, \cite{banerjee2} and \cite{yuanlin} independently proposed the $\ell_1$ norm penalized log-likelihood maximization approach, also known as Graphical Lasso or GLasso estimator. Methods that employ $\ell_1$ norm penalization include the Neighborhood Selection \citep{meinshausenbuhlman}, Sparse Partial Correlation Estimation \citep{peng}, constrained $\ell_1$ norm minimization for inverse matrix estimation or CLIME \citep{cai}, $\ell_1$ norm penalized D-trace loss minimization \citep{zhangzou}, Sparse Column-Wise Inverse Operator or SCIO \citep{liu2015fast}, among several others. Despite the popularity of $\ell_1$ norm penalty, other penalties (both convex and non-convex) are also considered, including  Adaptive $\ell_1$ norm \citep{fan, avagyan}, SCAD \citep{fan}, Ridge or $\ell_2$ norm \citep{van2016ridge, kuismin2017precision, bekker2023computational}, Elastic-Net \citep{kovacs2021graphical, bernardini2022new}, among others. We provide more details on estimation techniques in the next section.

In this article, we study the estimation of precision matrix $\Omega$ without assuming a specific structure or a sparsity pattern. In line with above-cited literature, our proposed method uses $\ell_1$ norm penalized log-likelihood optimization problem. Although this penalty produces a sparse estimate, it can also lead to an increased uncertainty in capturing the true relationships between variables. This is particularly important when Lasso incorrectly shrinks some of the entries to zero. Moreover, Lasso penalization tends to shrink the eigenvalues of the estimated precision matrix. In general, smaller eigenvalues of a precision matrix correspond to higher uncertainty in the data structure. In order to diminish the induced uncertainty, we consider an additional log-determinant penalty on our optimization problem. This term is known as an appropriate measure of uncertainty in a context of multivariate normal distribution and is directly related to entropy \citep{anderson, bishop2006pattern}. Our introduced method therefore induces sparsity through Lasso penalty term and adjusts uncertainty through the log-determinant term. The new augmented penalty is the convex combination of these two terms, motivated by the Elastic-Net penalization framework of \cite{kovacs2021graphical} and \cite{bernardini2022new}. 

We evaluate the performances of the proposed estimator through an extensive numerical study. In particular, for our simulation analyses we consider different patterns (i.e., models) of the true precision matrix $\Omega$, including those used in the experiments of the previous studies. We measure the performance of the considered methods in terms of statistical losses and GGM prediction measures. Finally, for the proposed method, we establish the convergence rate in the Frobenius norm under standard
asymptotic theory (i.e., we assume that the number of variables is fixed).

The manuscript is organized as follows. In Section \ref{section:2}, we describe the related work and our proposed methodology. In Section \ref{section:4}, we evaluate the statistical performance of the proposed methodology and compare it with that of other approaches in the literature. In Section \ref{section:5}, we provide real data applications: classification of prostate cancer patients using Linear Discriminant Analysis and computation of an optimal financial portfolio. We provide our conclusions in Section \ref{section:6}. We demonstrate technical details in Appendix A, simulation models in Appendix B and additional numerical results in Appendix C. Finally, Appendix D develops the analytical convergence rate of the proposed method. 

\section{Methodology}
\label{section:2}
\subsection{Background and existing research}
\label{section:2.1}

We use the following mathematical notations throughout this paper. For any symmetric matrix $\textbf{A}=[a_{ij}]_{1\le i,j\le p} \in \mathbb{R}^{p\times p}$, we denote the Frobenius  or $\ell_2$ norm by $||\textbf{A}||_2=\sqrt{\sum\limits_{i=1}^{p}\sum\limits_{j=1}^{p}a_{ij}^2}$, the $\ell_{\infty}$ norm by $||\textbf{A}||_{\infty}=\max\limits_{1\le i,j\le p}|a_{ij}|$, the matrix $\ell_1$ norm by $||\textbf{A}||_{\ell_1}=\max\limits_{1\le j\le p}\sum\limits_{i=1}^p|a_{ij}|$, the elementwise $\ell_1$ norm by $||\textbf{A}||_{1}=\sum\limits_{i=1}^{p}\sum\limits_{j=1}^{p}|a_{ij}|$, the spectral norm by $||\textbf{A}||_{\text{sp}}=\sup\limits_{||x||_2\le 1}||\textbf{A}x||_2$. Here, we define the $\ell_{2}$ norm by $||\textbf{a}||_2=\sqrt{\sum\limits_{j=1}^{p}a_j^2}$ for any $p$-dimensional vector $\textbf{a}=(a_1,...,a_p)^T\in \mathbb{R}^p$.

Without loss of generality, we assume that $\textbf{X}_{n\times p}$ is mean-centered, observed sample data matrix, where each row $X_{i} =\left(X_{i1},...,X_{ip}\right)$ is a $p$-dimensional normal random vector, i.i.d. for $i=1,...,n$ and has a covariance matrix $\Sigma=\Omega^{-1} \in \mathbb{R}^{p\times p}$. We define the sample covariance matrix as  $S=\dfrac{1}{n}\sum\limits_{i=1}^{n}X_i^{T}X_i  = \dfrac{1}{n}\textbf{X}^T\textbf{X}$.

The estimation of a precision matrix under high dimensional settings has been widely studied in prior scholarly work. Among the proposed methods, the well-explored estimator is based on minimizing penalized (or constrained) negative log-likelihood function of a multivariate normal distribution. In particular, Graphical Lasso or GLasso \citep{banerjee2, yuanlin, friedman} is the most popular approach for estimating the precision matrix and the corresponding GGM. This estimator is defined as:
\begin{eqnarray}
	\label{GLASSO}
	\centering
	\widehat{\Omega}_{\text{GLasso}}&=&\arg\min_{\Omega}\ -\log\det(\Omega)+\text{trace}(\Omega S)+\gamma||\Omega||_{1},
\end{eqnarray}
where the Lasso or $\ell_1$-norm penalty is used to encourage sparse solutions. Here, $\gamma>0$ is the associated penalty (i.e., tuning or calibration) parameter that determines the degree of applied penalization on $\hat{\Omega}$. In case of GLasso, this parameter adjusts the accuracy and the sparsity of precision matrix estimator. When $\gamma = 0$, the problem (\ref{GLASSO}) leads to the Maximum Likelihood Estimate of the covariance matrix $\hat{\Sigma} = S$.

Extant literature has studied the convergence rates of the GLasso precision matrix estimate $\widehat{\Omega}_{\text{GLasso}}$ under certain assumptions. In particular, \cite{rothman2} derived the convergence rate of the Glasso estimate in terms of Frobenius norm $\ell_2$, whereas \cite{ravikumar} established the convergence rate in terms of $\ell_\infty$ and spectral norms. Moreover, \cite{ravikumar} prove the model selection consistency, i.e., show the estimate $\widehat{\Omega}_{\text{Lasso}}$ specifies correctly the sparsity pattern of $\Omega$,  with probability converging to one.

In the context of penalization-driven approaches, a natural alternative to the Lasso penalization is the Ridge (or $\ell_2$-norm) penalization. The most straightforward version of this estimator is defined as:
\begin{eqnarray}
	\label{GRidge}
	\centering
		\widehat{\Omega}_{\text{GRidge}}&=&\arg\min_{\Omega}\ -\log\det(\Omega)+\text{trace}(\Omega S)+\gamma||\Omega||_{2}^2.
\end{eqnarray}
This estimator was initially discussed by \cite{witten2009covariance} and \cite{rothman2} and was further developed independently by \cite{van2016ridge} and \cite{kuismin2017precision}. \cite{kuismin2017precision} refer to it as ROPE (Ridge-type Operator for the Precision matrix Estimation). On the other hand, \cite{van2016ridge} suggested an alternative Ridge-type estimator, defined as: 
\begin{eqnarray}
	\label{GRidgeT}
	\centering
	\widehat{\Omega}_{\text{T-GRidge}}&=&\arg\min_{\Omega}\ -\log\det(\Omega)+\text{trace}(\Omega S)+\dfrac{\gamma}{2}||\Omega - {T}||_{2}^2,
\end{eqnarray}
where ${T} \in \mathbb{R}^{p\times p}$ is a symmetric positive definite target matrix. Clearly, the penalty term of $(\ref{GRidge})$ is the special case of $(\ref{GRidgeT})$ with ${T}=\textbf{0}$. Note that Ridge-type estimators are less complex than GLasso estimator (\ref{GRidge}) because they can be calculated by an explicit closed-form solutions. However, Ridge penalty does not induce exact sparsity on the estimated precision matrix, similar to Lasso penalty. In this way, these estimators are more attractive, if the sparsity of $\Omega$ is not an important property for the precision matrix estimate. For the sake of simplicity, analogously to the Graphical Lasso, we use the term Graphical Ridge (or GRidge) for the Ridge-type estimator $(\ref{GRidge})$ with no target, and Targeted Graphical Ridge (or T-GRidge) for the Ridge-type estimator $(\ref{GRidgeT})$ with a specified target. 

More recently, scholarly work has studied the convex combination of Lasso and Ridge penalties. In the regression context, this penalty is known as Elastic-Net \citep{zou2005regularization}. \cite{bernardini2022new} consider the following augmented estimation problem for estimating the precision matrix $\Omega$:
\begin{eqnarray}
	\label{ElNet}
	\centering
	\widehat{\Omega}_{\text{GEN}}&=&\arg\min_{\Omega}\ -\log\det(\Omega)+\text{trace}(\Omega S)+\gamma\left(\alpha ||\Omega||_{1} + (1-\alpha) ||\Omega||_{2}^2\right),
\end{eqnarray}
where the parameter $\alpha \in [0;1]$ controls the balance between Lasso and Ridge penalization. For $\alpha = 1$, we have the Graphical Lasso approach (\ref{GLASSO}), and for $\alpha = 0$, we have the Rope approach (\ref{GRidgeT}). Consistent with the terminology in regression context, this approach is called Graphical Elastic-Net (or GEN). The Elastic-Net penalization often brings the benefits of both Lasso and Ridge penalties: it leads to sparse estimators and simultaneously encourages stability in a presence of highly correlated variables.

Similar to the targeted Ridge penalization (\ref{GRidgeT}), \cite{kovacs2021graphical} suggested a targeted version for the Elastic-Net penalty:
\begin{eqnarray}
	\label{ElNet-T}
	\centering
	\widehat{\Omega}_{\text{T-GEN}}&=&\arg\min_{\Omega}\  -\log\det(\Omega)+\text{trace}(\Omega S) \nonumber\\
	&+&\gamma\left(\alpha ||\Omega - T||_{1} + \dfrac{1-\alpha}{2} ||\Omega - T||_{2}^2\right).
\end{eqnarray}
Several target matrices are suggested in \cite{kovacs2021graphical}. The penalty term of $(\ref{ElNet})$ is a special case of $(\ref{ElNet-T})$ with ${T}=\textbf{0}$, and the penalty term of $(\ref{GRidgeT})$ is a special case of $(\ref{ElNet-T})$ with $\alpha=0$.

Note that the popularity of $\ell_1$ and $\ell_2$ norm penalties is due to their convexity. The literature also considers applying nonconcave penalties on the log-likelihood function, such as Smoothly Clipped Absolute Deviation or SCAD \citep{fan}, $\ell_0$ norm penalty \citep{marjanovic2015l, liu2016sparse}, $\ell_q$ norm penalty with $0<q<1$ \citep{marjanovic2014}. For a further review on approaches for estimating high dimensional precision matrices, we refer to \cite{fan2016overview}.

\subsection{Proposed methodology}
\label{section:2.2}
Following the Elastic-Net-type penalization (\ref{ElNet}), we propose a new approach for estimating the precision matrix. Consider the following optimization problem:
\begin{eqnarray}
	\label{EAGL}
	\centering
	\widehat{\Omega}_{\text{EAGL}}&=&\arg\min_{\Omega}\  -\log\det(\Omega)+\text{trace}(\Omega S) \nonumber\\
	&+&\gamma\left( \alpha ||\Omega||_{1} + (1-\alpha) \log\det(\Omega^{-1}) \right),
\end{eqnarray}
where $\alpha\in (0;1)$ is a combination parameter analogous to its use in the Elastic-Net penalization (\ref{ElNet}). The proposed approach is an augmented version of the GLasso method (\ref{GLASSO}) with an additional $H_C(\textbf{X})=\log\det(\Omega^{-1})=\log(\det(\Sigma))$ penalty term. Here, we still use the Lasso penalty, because our goal is to obtain a sparse estimate. The motivation behind the $\log\det$ term is twofold. 

First, we can show that this term represents the entropy of multivariate Gaussian distribution, up to an additive constant. We show this in Appendix A. In the context of machine learning and statistics, entropy is viewed as a measure of uncertainty or randomness within a dataset. High entropy indicates more randomness, while low entropy suggests a more organized data and thus, better measurability of the relationships among the variables. 
The objective of the proposed technique is therefore to impose sparsity on the precision matrix estimate while adjusting the overall uncertainty, which leads to a more efficient estimate.

\begin{figure}[h!]
	
	\caption{Eigenvalues of GLasso (left) and EAGL (right) estimators for different value of the penalty parameter $\gamma$.}
	\label{Fig111}
	\includegraphics[width=80mm]{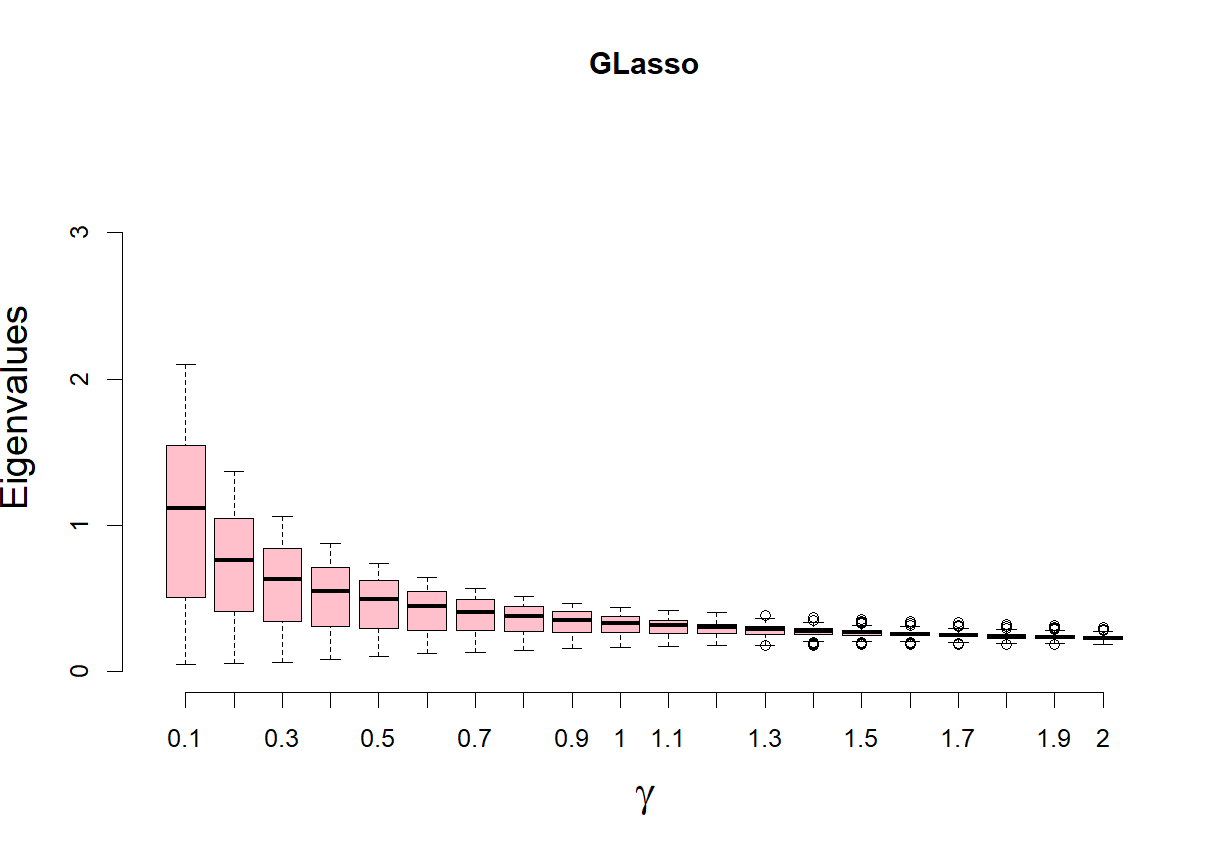} \ 
	\includegraphics[width=80mm]{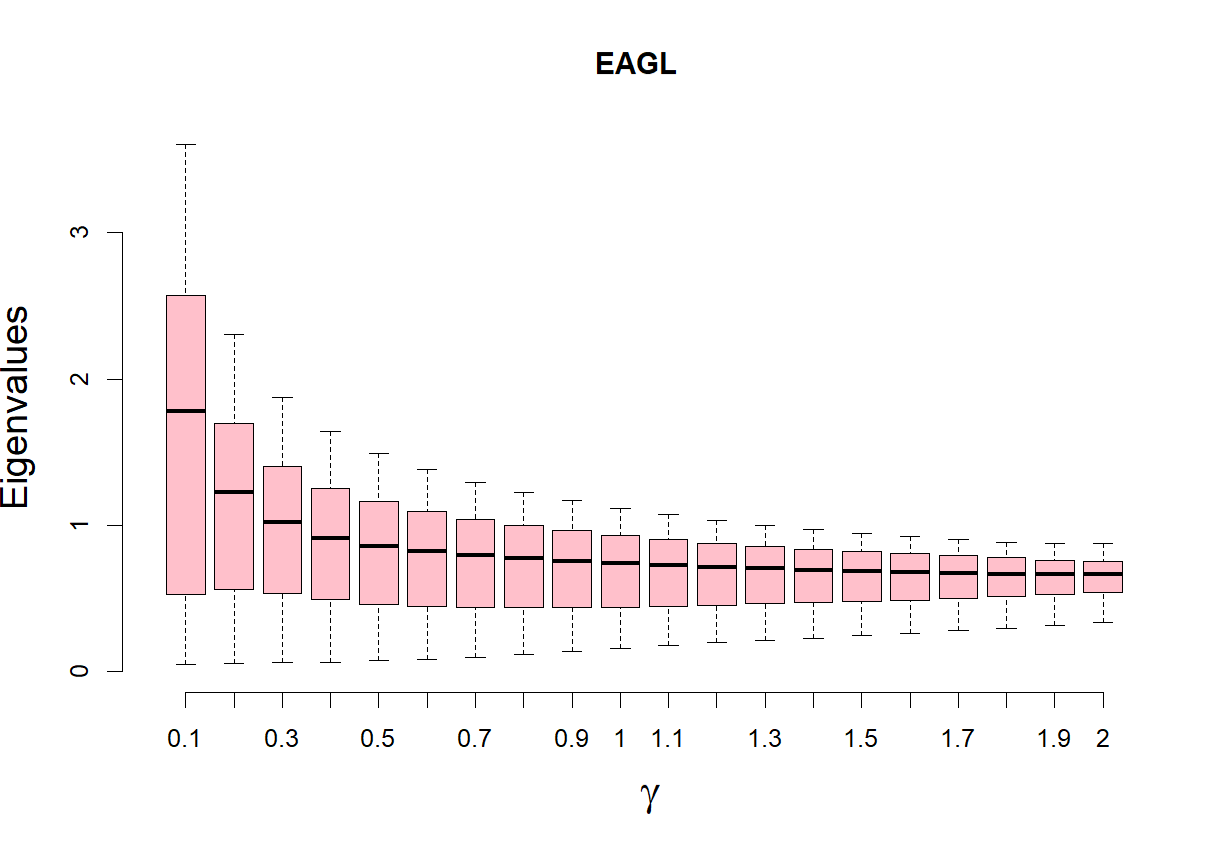}
\end{figure}

Second, the term $H_C(\textbf{X})$ produces an additional calibration for the eigenvalues of estimated precision matrix. Note that $H_C(\textbf{X})=\log\det(\Omega^{-1})=-\sum_{i=1}^{p}log(\lambda_{i})$, where $\lambda_{i}$ is the $i$-th eigenvalue of the precision matrix $\Omega$. The additional $\log\det$ penalty forces the precision matrix eigenvalues to shrink less which occurs due to the Lasso penalization (such shrinkage pattern is in line with the findings by  \cite{friedman} and \cite{avagyan2021d}). As a result, the eigenvalues of the estimated precision matrix $\widehat{\Omega}_{\text{EAGL}}$ become higher and less shrunk than those of $\widehat{\Omega}_{\text{GLasso}}$. In Figure \ref{Fig111}, we provide a simple example, which shows the distribution of eigenvalues of $\widehat{\Omega}_{\text{GLasso}}$ and $\widehat{\Omega}_{\text{EAGL}}$ for different values of the penalization parameter $\gamma$ from a sequence varying from 0.1 to 2. For the data simulation, we use the settings described in Section \ref{section:4} and we assume that the true precision matrix has a structure given in Model 1 for $p=200$ and $n=100$.  We can clearly see that the eigenvalues of $\widehat{\Omega}_{\text{EAGL}}$ shrink slower than those of $\widehat{\Omega}_{\text{GLasso}}$. Therefore, we can expect a more accurate eigenspectrum of the estimated precision matrix with the additional penalty term.

Note that large precision matrix eigenvalues indicate strong concentration in the principal axes and thus, lower uncertainty. This behavior aligns with the intuition that as the matrix becomes sparser due to the Lasso penalization, it reflects to fewer restrictions on the relationships between the variables (i.e., many variables become conditionally independent given the others). This will lead to a greater variability in the joint distribution (i.e., it will be more spread out) and, consequently, to a higher entropy.

The use of $\log\det$ penalty is not new in the literature for other purposes. For example, a similar $\log\det$ penalty is used by \cite{rothman2} for estimating sparse covariance matrix. However, this penalty was used as a logarithmic barrier to ensure that the obtained estimator is positive definite. To the best of our knowledge, this term has not been used in the context of entropy adjustment.

We call the proposed approach \textit{Entropy Adjusted Graphical Lasso} or EAGL. Note that when $\alpha=1$, the problem (\ref{EAGL}) reduces to (\ref{GLASSO}). Interestingly, when $\alpha = 0$, the problem (\ref{EAGL}) leads to the covariance matrix estimator $\hat{\Sigma} = \dfrac{1}{1 + \gamma} S$, which is a special case of the James-Stein estimator \citep{james1992estimation} and has a better performance than the traditional estimator $\hat{\Sigma} = S$ in terms of the MSE \citep[see also ][for a similar covariance estimator]{haff1980empirical}. This estimator can also be seen as the Ledoit-Wolf shrinkage covariance estimator \citep{ledoit2003honey} towards a zero target.   



\textbf{Remark}: Following the targeted version of the Elastic-Net estimator (\ref{ElNet}), we can also employ an additional target matrix $T$ in the Lasso penalty. However, in this paper we focus on the analysis of EAGL estimator without using target matrices.

\section{Simulation analysis}
\label{section:4}

In this section, we evaluate the numerical performance of the proposed method using simulated data. The data generation process is based on different scenarios (i.e., models) for the true precision matrix $\Omega$. Specifically, we consider several popular methods for estimating the precision matrix.  We compare our proposed estimator EAGL with Graphical Lasso,  Graphical Elastic-Net (with and without a target), Graphical Ridge (with and without a target) and Graphical SCAD. Our study is implemented in $\mathbf{R}$ using the following packages: \verb|glassoFast| for EAGL and GLasso, \verb|GLassoElnetFast| for Graphical Elastic-Net estimators, \verb|rope| for Graphical Ridge estimators, \verb|ggmncv| for Graphical SCAD. All packages are available at \url{http://cran.r-project.org/web/packages}. Following \cite{kuismin2017precision}, we use scalar matrix $T = \nu \textbf{I}$ for the target matrix, which is used in all targeted estimators (here, $\nu = p / \text{trace}(S)$ and $\textbf{I}\in \mathbb{R}^{p\times p}$ is identity matrix). As indicated earlier, Graphical Ridge approaches do not provide sparse estimators. Therefore, we perform an additional sparsification on these estimates using \verb|sparsify()| function of \verb|rags2ridges| package, with  `localFDR' thresholding argument \citep[see][for more details]{peeters2020rags2ridges}.

\subsection{Performance evaluation}
\label{section:4.1}
We evaluate the performance of a precision matrix estimator based on several popular statistical distances and errors. We consider the relative entropy losses, such as the Kullback-Leibler loss and the Reverse Kullback-Leibler loss \citep[see, for instance,][etc.]{yuan, yinli, avagyan2021d}, the Relative Trace Error (RTE) and matrix losses, such as the Frobenius $\ell_2$ norm, the spectral $\ell_{sp}$ norm and the matrix $\ell_1$ norm \citep[see, for instance,][etc.]{cai, zhangzou, van2016ridge}. These measures are defined as:
\begin{eqnarray*}
	\label{KLL}
	\text{KLL}(\widehat{\Omega},\Omega)&=&\text{trace}(\Omega^{-1} \widehat{\Omega})-\log\det(\Omega^{-1} \widehat{\Omega})-p,\\
	\label{RKLL}
	\text{RKLL}(\widehat{\Omega},\Omega)&=&\text{trace}(\Omega \widehat{\Omega}^{-1})-\log\det(\Omega \widehat{\Omega}^{-1})-p,\\
	\label{RTE}
	\text{RTE}(\widehat{\Omega},\Omega)&=&\left|1-\dfrac{\text{trace}(\widehat{\Omega})}{\text{trace}(\Omega)}\right|\\
	\label{FN}
	\ell_2(\widehat{\Omega},\Omega)&=&||\widehat{\Omega}-\Omega||_2,\\
	\label{LN}
	\ell_{\text{sp}}(\widehat{\Omega},\Omega)&=&||\widehat{\Omega}-\Omega||_{\text{sp}},\\
	\label{L1}
	\ell_1(\widehat{\Omega},\Omega)&=&||\widehat{\Omega}-\Omega||_{\ell_1}.
\end{eqnarray*}

Furthermore, we evaluate the graphical modeling (i.e., sparsity pattern prediction of the GGM) based on overall summary metrics: Matthews Correlation Coefficient \citep{matthews} and Balanced Accuracy \citep{brodersen2010balanced}. These metrics are respectively defined as:
\begin{eqnarray*}
	\label{MCC}
	\text{MCC}&=&\dfrac{\text{TP}\times \text{TN}-\text{FP}\times \text{FN}}{\sqrt{(\text{TP}+\text{FP})(\text{TP}+\text{FN})(\text{TN}+\text{FP})(\text{TN}+\text{FN})}},\\ 
	\label{BA}
	\vspace{0.7cm}
\text{BA}&=&\dfrac{1}{2} \left( \dfrac{TP}{TP + FN }  + \dfrac{TN}{TN + FP} \right).
\end{eqnarray*}

Here, TP is the number of correctly selected non-zero entries (i.e., true positives), TN is the number of correctly selected zero entries (i.e., true negatives), FP is the number of incorrectly selected non-zero entries (i.e., false positives), and FN is the number of incorrectly selected zero entries (i.e.,  false negatives). In case of MCC, the coefficient of 1 and $-1$ indicate the perfect and the worst classification, respectively. In case of the Balanced Accuracy, the coefficient 1 indicates the perfect selection, whereas the coefficient $0$ indicates the worst selection. For convenience, we further rescale the MCC to the range of the Balanced Accuracy.
\begin{eqnarray*}
	\label{uMCC}
	\text{uMCC}&=&\dfrac{\text{MCC}+1}{2}.
\end{eqnarray*}
Both metrics are popular techniques in statistics and machine learning as standard accuracy metrics for evaluating binary classifications \citep[see ][ for more details]{chicco}. 



\subsection{Simulation settings}
\label{section:4.2}
We generate multivariate normal random samples with zero mean and covariance matrix $\Sigma=\Omega^{-1}$. The following models are considered for the true precision matrix $\Omega = [\omega_{ij}]_{1\le i, j \le p}$:
\begin{itemize}
	\item \textbf{Model} 1. Band matrix, with $\omega_{ii}=1$, $\omega_{i,i-1}=\omega_{i-1,i}=0.45$ and other values are 0 \citep[][etc.]{yuanlin, friedman}
	\item \textbf{Model} 2. Band matrix, with $\omega_{ii}=1$, $\omega_{i,i-1}=\omega_{i-1,i}=0.5$, $\omega_{i,i-2}=\omega_{i-2,i}=0.35$ and other values are 0  \citep[][]{kuismin2017precision, avagyan2021d}.
	\item \textbf{Model} 3. A block-diagonal matrix, with four equally sized blocks along the
	diagonal. Each block is defined as $\omega_{ij}=0.6^{|i-j|}$ \citep[][etc.]{cai, fan}
	\item \textbf{Model} 4. A random positive definite matrix, with approximately $50\%$ of non-zero entries. This matrix is generated using Matlab command \texttt{sprandsym} with a parameter $0.5$ \citep{avagyan2021d}. 
	\item \textbf{Model} 5. Erd\H{o}s-R\'{e}nyi random graph. We define $\Theta = [\theta_{ij}]_{1\le i, j \le p}$, with $\theta_{ij} = u_{ij}  \delta_{ij}$, where $u_{ij} = u_{ji}$ is a uniform random variable on an interval $[0.4, 0.8]$ and $\delta_{ij}=\delta_{ji}$ is a Bernoulli random variable with a success probability $0.05$. Next, we take $\Omega = \Theta + (|\lambda_{\text{min}}(\Theta) + 0.05|) \textbf{I}$, where $\lambda_{\text{min}}(\Theta)$ is the smallest eigenvalue of $\Theta$ \citep{kovacs2021graphical}.
	\item \textbf{Model} 6. Scale-free graph, based on the Barab\'asi–Albert algorithm. The resulting graph has $p$ edges. This matrix is generated using R package \texttt{huge} \citep{kovacs2021graphical, bernardini2022new}. 
	\item \textbf{Model} 7. Hub graph with 10 hubs, each hub containing $p/20$ nodes. The resulting graph has $p-10$ edges. This matrix is generated using R package \texttt{huge} with parameters $\nu=0.3$ and $u=0.1$ \citep{kovacs2021graphical, bernardini2022new}.
\end{itemize}

The generated matrices are further standardized to have unit diagonal. We provide the graphical models corresponding to each considered precision matrix structure in Appendix B. Finally, we set the sample size $n=100$ and the number of variables $p=200$. The number of replications is 100.

\subsection{Penalty parameter selection}
\label{section:4.3}
Selecting the penalty parameter is a crucial step for any penalized optimization problem. In our numerical analysis, we select the penalty parameter $\gamma$ using 5-fold Cross-Validation or CV \citep{bien2011sparse} approach. This is a commonly used technique in precision matrix estimation methods and is consistent with studies of \cite{kovacs2021graphical}, \cite{kuismin2017precision}, etc.  Following \cite{bernardini2022new} and \cite{kovacs2021graphical}, we set $\alpha = 0.5$ for the estimators GEN, T-GEN and the proposed EAGL.

Another popular technique for estimating the penalty parameter of the precision matrix estimation methods is Bayesian Information Criterion or BIC \citep {yuanlin}. In this paper, we put CV above BIC, because CV leads to lower statistical losses. However, according to \cite{wasserman}, CV may provide many false positives compared to BIC. 
Following \cite{bernardini2022new}, we conduct additional analyses for demonstrating the performance of our proposed methodology where the penalty parameter $\gamma$ is selected using BIC approach. We provide these results in Appendix C. These findings provide further insights into the comparison of the two selection techniques.

\subsection{Computational time}
\label{section:4.4}
Table \ref{table:time} shows the average computational time required for each method. The reported time includes the required time to select the penalty parameter $\gamma$ using CV and the required time to run the method. We assume that the true precision matrix has a structure given in Model 1 for $p=200$ and $n=100$. The analyses are performed on Intel Core i7-1280P CPU, 1.80 GHz. The computational time of EAGL is relatively low, compared to the other methods. Note that, in general, Ridge-type approaches are not complex and require little computation (usually less than a second). However, the time provided in Table \ref{table:time} is high due to the sparsification procedure.

\renewcommand\arraystretch{0.8}
\begin{table}[ht!]
	\caption{Average computational time (in seconds) for different methods.}
	\vspace{0.5cm}
	\label{table:time}
	\begin{center}
		\resizebox{14cm}{!}{
			\begin{tabular}{@{\extracolsep{0.5cm}}c@{\extracolsep{0.5cm}}c@{\extracolsep{0.5cm}}c@{\extracolsep{0.5cm}}c@{\extracolsep{0.5cm}}c@{\extracolsep{0.5cm}}c@{\extracolsep{0.5cm}}c@{\extracolsep{0.5cm}}c}
				\hline\hline
				
				$\widehat{\Omega}_\text{GLasso}$  &  $\widehat{\Omega}_\text{GEN}$ & $\widehat{\Omega}_\text{T-GEN}$ & $\widehat{\Omega}_\text{GRidge}$
				& $\widehat{\Omega}_\text{T-GRidge}$  & $\widehat{\Omega}_\text{SCAD}$  & $\widehat{\Omega}_\text{EAGL}$   \\
				\hline
				 2.88 &	5.44 &	4.36    & 13.1 & 12.6 &	2.73 &	1.86  \\\hline
				
				\hline
				\hline
			\end{tabular}
		}
	\end{center}
\end{table}

\subsection{Discussion of results}
\label{section:4.5}

Tables \ref{table:Summary01}-\ref{table:Summary07} report the averages of the measures over 100 replications. Corresponding standard deviations are provided in the parentheses.  First, we observe that EAGL outperforms GLasso in terms of RKLL, RTE, $\ell_2$, $\ell_{\text{sp}}$ and $\ell_1$ norms for all models, in terms of KLL for models 1, 3, 6 and 7, and in terms of uMCC for models 1, 2, 3, and 7. On the other hand, GLasso method performs better than EAGL in terms of KLL for models 2, 4 and 5 and in terms of uMCC for models 4, 5 and 6. Table \ref{table:Summary001} provides a summary for the comparison between EAGL and GLasso estimators, indicating scenarios, where EAGL outperforms GLasso.

\begin{center}
	\textbf{Tables \ref{table:Summary01} - \ref{table:Summary07} about here.}
\end{center}

Comparing our proposed method with other methods, we see that in general EAGL provides better results, especially in terms of the statistical losses. However, SCAD provides the best overall results in terms of the KLL for models 1, 2, 3, 4, in terms of the $\ell_1$ norm for model 7 and in terms of uMCC for model 5. We observe that T-GEN estimator provides the best overall results in terms of KLL for models 5 and 6, in terms of $\ell_2$ norm in terms for model 6, in terms of the $\ell_1$ norm for model 6 and in terms of BA for model 4 and 5. On the other hand, GEN estimator provides the best overall results in terms of  BA for model 6. T-GRidge estimator provides the best overall results in terms of the $\ell_1$ norm for model 3, in terms of BA for model 1 and in terms of uMCC for model 7. Finally, GRidge estimator provides the best overall results in terms of the $\ell_1$ norm for models 1 and 2, in terms of  uMCC for models 1, 2, 3 and 6. 

\renewcommand\arraystretch{1.2}

\begin{table}[ht!]
	\caption{EAGL (\checkmark) vs GLasso (\text{\sffamily X}): A Comparative Overview.}.
	\label{table:Summary001}
	\begin{center}
		\resizebox{7cm}{!}{
			\begin{tabular}{c|@{\extracolsep{0.5cm}}c@{\extracolsep{0.5cm}}c@{\extracolsep{0.5cm}}c@{\extracolsep{0.5cm}}c@{\extracolsep{0.5cm}}c@{\extracolsep{0.5cm}}c@{\extracolsep{0.5cm}}c}
				\hline\hline
				
				Metrics &   \multicolumn{7}{c}{Models}     \\	\hline\hline
				&  1  &  2  &  3 &  4	&  5  &  6  &  7   \\
				\hline
				KLL		& \checkmark &\text{\sffamily X}&	\checkmark & \text{\sffamily X} & \text{\sffamily X} & \checkmark &\checkmark \\\hline
				
				RKLL	& \checkmark &	\checkmark &	\checkmark & \checkmark  &	\checkmark & \checkmark &\checkmark  \\\hline
				
				$\ell_2$&\checkmark &	\checkmark &	\checkmark & \checkmark  &	\checkmark & \checkmark &\checkmark  \\\hline
				
				$\ell_{\text{sp}}$&\checkmark &	\checkmark &	\checkmark & \checkmark  &	\checkmark & \checkmark &\checkmark \\\hline
				
				$\ell_1$	&\checkmark &	\checkmark &	\checkmark & \checkmark  &	\checkmark & \checkmark &\checkmark \\\hline
				
				RTE			&\checkmark &	\checkmark &	\checkmark & \checkmark  &	\checkmark & \checkmark &\checkmark \\\hline
				
				uMCC		& \checkmark &	\checkmark &	\checkmark & \checkmark  &	\checkmark & \checkmark &\checkmark 
				
				\\\hline
				
				BA			&\checkmark &	\checkmark &	\checkmark & \text{\sffamily X} & \text{\sffamily X} & \text{\sffamily X} &\checkmark \\\hline

				\hline
				\hline
			\end{tabular}
		}
	\end{center}
\end{table}

In sum, the proposed EAGL estimator in general provides favorable performance than GLASSO, GEN, T-GEN, GRidge, T-GRidge and SCAD methods for most of the models in terms of most statistical losses and GGM prediction measures. Moreover, EAGL shows a suitable trade-off between the matrix prediction and sparsity pattern identification, i.e., the outperformance in terms of one criterion does not diminish the other one. This suggests that entropy adjustment enhances both the statistical accuracy of the estimated precision matrix and the performance of GGM selection.

In addition, we conduct the same study where the penalty parameter is selected using BIC (see Appendix C). The results support our discussion above and the comparison of the considered methods remains comparatively the same.

\section{Real data application}
\label{section:5}
In this section, we conduct an empirical analysis of the proposed method through two real-data applications. The first application aims at predicting prostate cancer patients and the 
second one aims at selecting a large financial portfolio.

\subsection{Prostate cancer study}
\label{section:5.1}
In this application, we focus on the problem of predicting prostate cancer patients using Linear Discriminant Analysis (LDA) with different precision matrix estimates. We use a dataset analysed by \cite{singh2002gene}, which contains measurements of the gene expression levels of 6033 genes for 102 specimens. Out of these specimens, 52 are taken from prostate cancer patients, and 50 are taken from healthy patients. The dataset \verb|singh2002| is available in R package \verb|sda|.

In order to evaluate the performance of the proposed EAGL method, we first divide the data into a training set and a testing set with sizes 60 and 42, respectively.  Second, for the training set we apply two sample t-test between the two groups (cancer and healthy) in order to select the most significant 100 genes with the smallest p-values. Based on the selected genes, we obtain the estimated precision matrix $\hat{\Omega}$ using the methods considered in section \ref{section:4}. The penalty parameters are estimated using the 5-fold CV technique. The estimated precision matrix is then used in the following LDA procedure. 

Let's assume $\boldsymbol\mu_1$ and $\boldsymbol\mu_2$ are the population means of the gene expression levels for cancer and healthy patients, respectively, and $\Omega$ is the corresponding population precision matrix. We use Mahalanobis distance $\textbf{a}^T(\textbf{x}-\boldsymbol\mu)$, where $\textbf{a} = \Omega (\boldsymbol\mu_1 - \boldsymbol\mu_2)$ and $\boldsymbol\mu = \dfrac{\boldsymbol\mu_1+\boldsymbol\mu_2}{2}$. We estimate the means $\boldsymbol\mu_1$ and $\boldsymbol\mu_2$ using within group averages $\bar{\textbf{x}}_1$ and $\bar{\textbf{x}}_2$, respectively, calculated with the training date. Next, we assign the testing observation $\textbf{x}$ to the cancer group if  $\hat{\textbf{a}}^T(\textbf{x}-\hat{\boldsymbol\mu})> 0$, where $\hat{\textbf{a}} = \hat{\Omega} (\bar{\textbf{x}}_1 - \bar{\textbf{x}}_2)$ and $\hat{\boldsymbol\mu} = \dfrac{\bar{\textbf{x}}_1+\bar{\textbf{x}}_2}{2}$. We repeat this process 100 times and calculate the missclassification rate for each precision matrix estimator. In addition, we calculate the uMCC and BA scores, where we assume that TP and TN are the number of correctly predicted cancer and healthy patients, respectively, and FP and FN are the number of incorrectly classified cancer and healthy patients, respectively.

Table \ref{table:class} reports the average prediction measures for each method. We observe that the proposed EAGL shows lower missclassification rate and higher uMCC and BA than the GLasso, GEN, GRidge, T-Gridg and SCAD estimators. We observe that EAGL and the targeted T-GEN have roughly the same classification measures. However, we note that in contrast to the T-GEN approach, the proposed EAGL approach does not require selecting a target matrix and is computationally more efficient.

\begin{table}[ht!]
	\caption{Average classification measures over 100 replications.}
	\vspace{0.5cm}
	\label{table:class}
	\begin{center}
		\resizebox{14cm}{!}{
			\begin{tabular}{@{\extracolsep{0.5cm}}c|@{\extracolsep{0.5cm}}c@{\extracolsep{0.5cm}}c@{\extracolsep{0.5cm}}c@{\extracolsep{0.5cm}}c@{\extracolsep{0.5cm}}c@{\extracolsep{0.5cm}}c@{\extracolsep{0.5cm}}c}
				\hline\hline
				
				& $\widehat{\Omega}_\text{GLasso}$  &  $\widehat{\Omega}_\text{GEN}$ & $\widehat{\Omega}_\text{T-GEN}$ & $\widehat{\Omega}_\text{GRidge}$
				& $\widehat{\Omega}_\text{T-GRidge}$  & $\widehat{\Omega}_\text{SCAD}$  &  $\widehat{\Omega}_\text{EAGL}$  \\
				\hline
				Missclassification rate		& 0.168	&	0.178	&	0.123	&	0.155	&	0.149	&	0.134	&	0.128	 \\ \hline
				uMCC						& 0.834	&	0.825	&	0.879	&	0.848	&	0.854	&	0.868	&	0.873	\\ \hline
				BA							& 0.831	&	0.822	&	0.876	&	0.845	&	0.851	&	0.865	&	0.871	 \\\hline
				\hline
			\end{tabular}
		}
	\end{center}
\end{table}

\subsection{S\&P 500 portfolio optimization}
\label{section:5.2}

In our second application, we focus on developing an optimal stock portfolio with minimum risk \citep{goto, avagyan2022precision, bernardini2022new}. Note that estimating the precision matrix has a crucial role in computing optimal portfolio weights \citep[see][for more details]{stevens}.  In the mean-variance optimization, the risk of a $p$-dimensional (global) portfolio $w=\left(w_1, ..., w_p\right)$ is measured by the standard deviation of its returns, given as $\sqrt{w^T\Sigma w}$ \citep{markowitz1952}, where $\Sigma$ is the global covariance matrix. In the absence of short-sale constraints, the estimated weights $\hat{w}_{MVP}$ of the single-period minimum variance portfolio are defined as:
\begin{eqnarray}
	\hat{w}_{MVP} &=& \arg \min_w w^T\hat{\Sigma} w \\
	& &\text{subject to } w^T\textbf{1} = 1 , \nonumber
\end{eqnarray}
\noindent where \textbf{1} is a $p$-dimensional vector of ones. \cite{demiguel} demonstrate that the minimum variance optimization problem has an explicit solution, defined as:
\begin{eqnarray}
	\hat{w}_{MVP} &=& \dfrac{\hat{\Sigma}^{-1}\textbf{1}}{\textbf{1}^T\hat{\Sigma}^{-1}\textbf{1}}.
\end{eqnarray}
This means that the minimum variance portfolio depends on the precision matrix estimate $\hat{\Omega} = \hat{\Sigma}^{-1}$. Therefore, we expect that an accurate estimate of the precision matrix will lead to a better portfolio.

We use historical monthly returns of $p=267$ stock constituents of S\&P 500 index for a total of $n = 364$ months, covering the period from January 1991 to April 2021. The constituents come from 11 major market sectors. The dataset is available at the \textbf{R} package \verb|probstats4econ|. Following \cite{ledoit2003honey}, we assume that stock returns are independent and identically distributed. In order to evaluate the performance of selected portfolio, we perform the `rolling-horizon' procedure of \cite{demiguel}. First, we select a rolling window with length $m=150$ months, which is used to estimate the precision matrix $\Omega$ and the corresponding portfolio weights. Next, we compute the out-of-sample portfolio returns of the next period. This procedure is repeated $n-m$ times by including the returns of next period and dropping the earliest one (thus, keeping the rolling window with the same size). Finally, we evaluate the performance of the portfolio by computing the out-of-sample mean (i.e., reward), the standard deviation (i.e., risk) and the Sharpe Ratio, based on the obtained $n-m$ out-of-sample returns $\hat{R}_t$:
\begin{eqnarray}
	\hat{\mu} &=& \frac{1}{n-m}\sum_{t = m}^{n-1}\hat{R}_{t+1}, \\
	\hat{\sigma}^2 & =& \frac{1}{n-m-1}\sum_{t = m}^{n-1}(\hat{R}_{t+1}-\hat{\mu})^2, \\
	\widehat{\text{SR}} &=& \frac{\hat{\mu}}{\hat{\sigma}} .
\end{eqnarray}
The Sharpe Ratio, also known as the reward-to-risk ratio, measures the performance of a portfolio relative to its associated risk. A higher Sharpe Ratio indicates better investment performance in terms of risk-adjusted returns.

In our analysis, in addition to the selected methods, we also evaluate the performance of two alternatives, commonly used in the literature. The first one is the inverse of Ledoit-Wolf covariance estimator $\hat{\Omega}_{\text{LW}} = \hat{\Sigma}_{\text{LW}}^{-1}$ \citep{ledoit2003honey}, which is a popular approach for selecting an optimal financial portfolio. The second one is the naive approach $\hat{\Omega}_{\text{Naive}} = \textbf{I}$, which leads to allocating $1/p$ weight of wealth to each of the $p$ assets available for investment \citep{demiguel2009optimal}. 

Table \ref{table:portfolio} presents the out-of-sample performances for portfolios obtained using different precision matrix estimates. The penalty parameter $\gamma$ is selected using 5-fold CV technique. First, our results show that the proposed EAGL method provides the lowest out-of-sample portfolio risk. EAGL, GLASSO and LW estimators lead to the highest out-of-sample mean (i.e., reward) compared to all the other methods. Finally, our proposed approach demonstrate the highest Sharpe Ratio. This means that the portfolios obtained using the proposed precision matrix estimate are the most efficient in terms of the risk-adjusted return. The results show that the naive approach provides the lowest out-of-sample mean and the highest risk. Finally, despite  its popularity, the LW estimator leads to relatively high risk.

\begin{table}[ht!]
	\caption{Out-of-sample performance (rounded) for different estimation methods.}
	\vspace{0.5cm}
	\label{table:portfolio}
	\begin{center}
		\resizebox{14cm}{!}{
			\begin{tabular}{@{\extracolsep{0.5cm}}c|@{\extracolsep{0.5cm}}c@{\extracolsep{0.5cm}}c@{\extracolsep{0.5cm}}c@{\extracolsep{0.5cm}}c@{\extracolsep{0.5cm}}c@{\extracolsep{0.5cm}}c@{\extracolsep{0.5cm}}c@{\extracolsep{0.5cm}}c@{\extracolsep{0.5cm}}c}
				\hline\hline
				
				& $\widehat{\Omega}_\text{GLasso}$  &  $\widehat{\Omega}_\text{GEN}$ & $\widehat{\Omega}_\text{T-GEN}$ & $\widehat{\Omega}_\text{GRidge}$
				& $\widehat{\Omega}_\text{T-GRidge}$  & $\widehat{\Omega}_\text{SCAD}$   & $\widehat{\Omega}_\text{LW}$  & $\widehat{\Omega}_\text{Naive}$ & $\widehat{\Omega}_\text{EAGL}$  \\
				\hline
				Mean		& 0.015 &	0.014	& 0.014		& 0.014 &	0.012  &	0.013 	&  0.015 &	0.012  &	0.015	 \\ \hline
				Risk		& 0.036  &	0.036	& 0.038		& 0.038	 & 	0.041  &	0.037 	&  0.039  &	0.045  & 	0.034\\ \hline
				Sharpe Ratio& 0.420  &	0.400  	& 0.386		& 0.385  &	0.293  &	0.364 	&  0.383  &	0.279  &	0.438	 \\\hline
				\hline
			\end{tabular}
		}
	\end{center}
\end{table}

\begin{figure}[h!]
	
	\caption{Graphical model representing the interactions between the companies from different sectors. To avoid complexity, the week partial correlations with $|\rho_{ij}| \le 2sd(NZ)$ are set to zero, where $NZ$ is the vector that contains the non-zero partial correlation entries.  }
	\vspace*{0.5cm}
	\label{Fig4}
	\begin{center}
		\includegraphics[width=160mm]{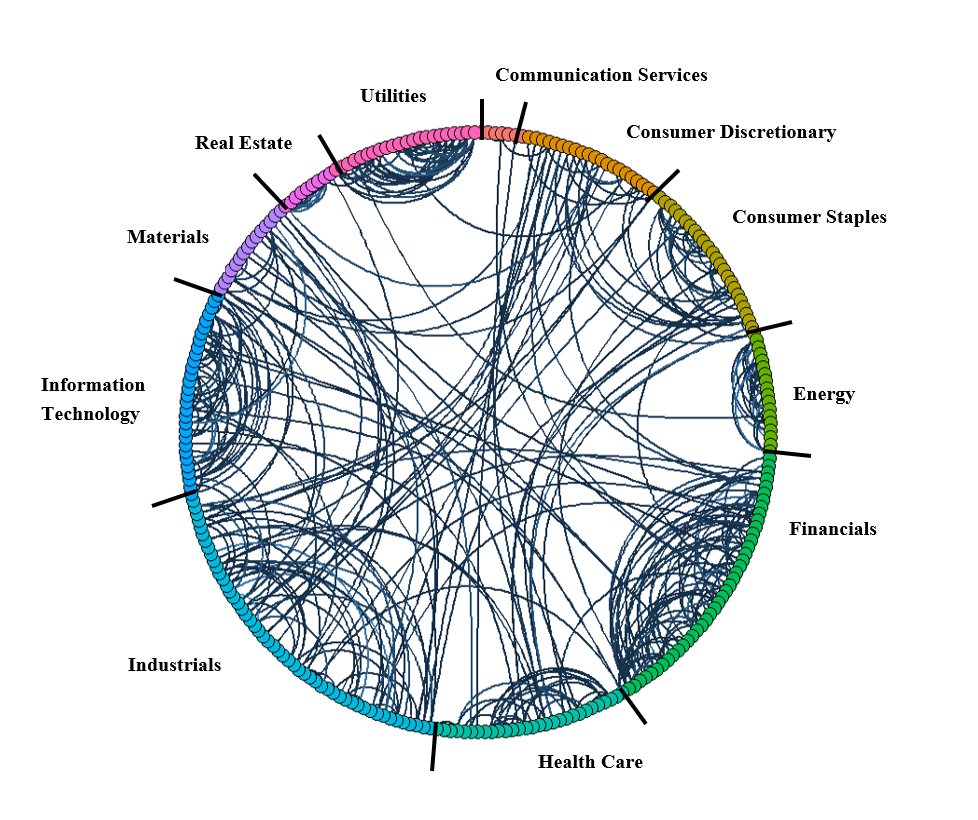} \\
		
	\end{center}
\end{figure}

In addition to the out-of-sample performance above, we also examine the interaction structure between the companies. For demonstrative purposes, we apply our proposed EAGL approach on the whole data. We observe that the partial correlation matrix (not shown), corresponding to the precision matrix, is roughly block-diagonal, where each block represents a market sectors. The interactions (i.e., partial correlations) are stronger among the companies within each sector. This indicates that there is a strong community structure, with communities made up of companies from the same sector. Figure \ref{Fig4} demonstrates the Gaussian Graphical Model illustrating the interactions among the companies. The GGM reveals groups of nodes (i.e., companies) such that within group interactions are stronger and more frequent than between group interactions.
\begin{center}
	\textbf{Figure \ref{Fig4} about here.}
\end{center}


\section{Conclusions}
\label{section:6}

 In this article, we introduce a novel method for estimating sparse precision matrices in high dimensional settings. The proposed approach is an augmented version of the popular Graphical Lasso method, which incorporates Lasso penalization framework with an additional log-determinant penalty. This combined penalty allows for entropy adjustment of the multivariate Gaussian distribution, thereby reducing the uncertainty.

Through extensive numerical analyses, using both simulated and real datasets, we demonstrate that the proposed entropy adjustment helps achieve better performance of the estimated precision matrix without increasing the computational costs. We evaluate our method using different loss functions and prediction performance measures. Our method provides lower statistical losses and better model selection metrics for Gaussian Graphical Models compared to other established approaches in the literature. We calibrate the penalty parameter using 5-fold Cross-Validation and BIC, ensuring practical usability across different datasets. Our proposed approach does not rely on the selection of a target matrix, though it can be extended to include a target matrix in the Lasso norm. Finally, we establish the convergence rate of the proposed EAGL estimator in the Frobenius norm under standard asymptotic conditions.

For researchers, we suggest incorporating the proposed EAGL in situations, where uncertainty reduction and sparsity are of high importance. This recommendation is particularly pertinent for scholars and practitioners who employ the precision matrix in finance, discriminant analysis and network modeling, because the improved accuracy of the estimated precision matrix can significantly improve their analyses. 

While our methodology has notable advantages, it is important to acknowledge certain limitations. First, in our study, we choose $\alpha=0.5$, i.e., halfway between Lasso and $\log\det$ penalties. However, the optimal value for $\alpha$ may differ from this choice. Second, $\alpha$ should be sufficiently far from zero, because EAGL estimate does not exist for $\alpha=0$, when $p$ is close to or exceeds $n$. Finally, we may encounter some convergence problems for EAGL for a relatively large norm of the true covariance matrix. 

Our results pave the way for future research. First, future research should further investigate the role of parameter $\alpha$ on numerical and theoretical performance of our methodology. Second, it might be interesting to explore the relationship between the parameters $\gamma$ and $\alpha$ in greater detail. Third, a promising research direction involves applying the proposed entropy adjustment to other precision matrix estimators, such as the Entropy Adjusted Graphical Ridge. Finally, studying the proposed estimator from a Bayesian perspective could provide valuable insights and deepen our understanding of its theoretical properties.

Despite the limitations, we hope our paper contributes valuable insights into precision matrix estimation and simulates future research in this area.

%
%

\newpage

\renewcommand\arraystretch{1.2}

\begin{table}[ht!]
	\caption{Average measures (with standard deviations) over 100 replications for Model 1. \textbf{Bold} letters indicate the best results. }
	\vspace{0.5cm}
	\label{table:Summary01}
	\begin{center}
		\resizebox{16cm}{!}{
			\begin{tabular}{c|@{\extracolsep{0.5cm}}c@{\extracolsep{0.5cm}}c@{\extracolsep{0.5cm}}c@{\extracolsep{0.5cm}}c@{\extracolsep{0.5cm}}c@{\extracolsep{0.5cm}}c@{\extracolsep{0.5cm}}c}
				\hline\hline
				
				& 	 $\widehat{\Omega}_\text{GLasso}$  &  $\widehat{\Omega}_\text{GEN}$ & $\widehat{\Omega}_\text{T-GEN}$ & $\widehat{\Omega}_\text{GRidge}$
				& $\widehat{\Omega}_\text{T-GRidge}$  & $\widehat{\Omega}_\text{SCAD}$  & $\widehat{\Omega}_\text{EAGL}$   \\
				\hline
				KLL		& 27.56 	(0.930) &	33.08	(0.627) & 30.86	(0.541) & 58.14	(1.399) & 64.03	(2.624) & \textbf{21.00 (0.591)} & 23.61	(0.839)  \\\hline
				
				RKLL	&  38.30	(2.767) &  49.72	(1.681) & 45.20 (0.918) & 75.97	(0.832) & 87.32	(5.451) & 25.49	(0.745) & \textbf{22.77	(0.845)} \\\hline
				
				$\ell_2$& 7.493	(0.401) &	8.365	(0.191) & 	8.193	(0.047)  	&	9.994	(0.019) & 10.40	(0.201) & 6.318	(0.078) & \textbf{6.107	(0.153)} \\\hline
				
				$\ell_{\text{sp}}$	& 1.041	(0.039) & 1.126	 (0.019)	& 1.105	(0.013) & 1.253	(0.010) & 1.308	(0.025) & 0.927	(0.023) &	\textbf{0.891	(0.023)} \\\hline
				
				$\ell_1$	& 1.654	(0.084) & 1.775	(0.066) &	1.688	(0.053) & \textbf{1.356	(0.026)} &	1.395	(0.029) &	1.467	(0.068) &	1.403	(0.068) \\\hline
				
				RTE		& 0.369	(0.030) &	0.419	(0.015) & 0.410	(0.003) &	0.482	(0.001) &	0.495	(0.015) &	0.297	(0.005) &	\textbf{0.256	(0.010)} \\\hline
				
				uMCC	& 0.653	(0.011) &	0.634	(0.002) &	0.648	(0.002) & \textbf{0.979	(0.007)} &	0.955	(0.018) &	0.681	(0.003) &	0.704	(0.004) \\\hline
				
				BA		& 0.936	(0.009) &	0.919	(0.002) &	0.933	(0.002) & 0.990	(0.006) & \textbf{0.990	(0.008)} & 0.955	(0.002) & 0.965	(0.001) 	 \\\hline

				\hline
				\hline
			\end{tabular}
		}
	\end{center}
\end{table}

\renewcommand\arraystretch{1.2}

\begin{table}[ht!]
	\caption{Average measures (with standard deviations) over 100 replications for Model 2. \textbf{Bold} letters indicate the best results. }
	\vspace{0.5cm}
	\label{table:Summary02}
	\begin{center}
		\resizebox{16cm}{!}{
			\begin{tabular}{c|@{\extracolsep{0.5cm}}c@{\extracolsep{0.5cm}}c@{\extracolsep{0.5cm}}c@{\extracolsep{0.5cm}}c@{\extracolsep{0.5cm}}c@{\extracolsep{0.5cm}}c@{\extracolsep{0.5cm}}c}
				\hline\hline
				
					& $\widehat{\Omega}_\text{GLasso}$  &  $\widehat{\Omega}_\text{GEN}$ & $\widehat{\Omega}_\text{T-GEN}$ & $\widehat{\Omega}_\text{GRidge}$
				& $\widehat{\Omega}_\text{T-GRidge}$  & $\widehat{\Omega}_\text{SCAD}$  & $\widehat{\Omega}_\text{EAGL}$   \\
				\hline
				KLL		&45.05	(1.137) &	47.27	(0.629) &	46.80	(0.706) &	69.07	(1.931) &	77.36	(0.849) &	\textbf{42.08	(1.267)} &	46.07	(1.135)  \\\hline
				
				RKLL	& 77.94	(6.640) &	84.77	(0.981) &	85.06	(3.230) &	106.7	(3.601) &	162.0	(2.687) &	63.59	(5.960) &\textbf{58.75	(1.691)} \\\hline
				
				$\ell_2$&	11.92	(0.393) &	12.17	(0.029) &	12.29	(0.180) &	13.15	(0.085) &	14.31	(0.040) &	11.27	(0.374) &	\textbf{11.21	(0.147)}\\\hline
				
				$\ell_{\text{sp}}$	&1.954	(0.054) &	1.984	(0.010) &	2.002	(0.025) &	2.096	(0.014) &	2.247	(0.006) &	1.875	(0.054) &	\textbf{1.867	(0.023)} \\\hline
				
				$\ell_1$	& 2.683	(0.067) &	2.773	(0.043) &	2.695	(0.050) &	\textbf{2.250	(0.027)} &	2.325	(0.014) &	2.602	(0.058) &	2.547	(0.060) \\\hline
				
				RTE		&0.467	(0.028) &	0.484	(0.002) &	0.489	(0.013) &	0.493	(0.007) &	0.573	(0.005) &	0.412	(0.026) &	\textbf{0.366	(0.010)}\\\hline
				
				uMCC	&0.671	(0.011) &	0.651	(0.002) &	0.668	(0.006) &	\textbf{0.873	(0.008)} &	0.827	(0.007) &	0.689	(0.009) &	0.703	(0.007) \\\hline
				
				BA		& 0.919	(0.010) &	0.900	(0.002) &	0.917	(0.005) &	0.829	(0.016) &	0.795	(0.014) &	0.933	(0.005) &	\textbf{0.940	(0.004)} \\\hline

				\hline
				\hline
				
			\end{tabular}
		}
	\end{center}

\end{table}

\newpage

\renewcommand\arraystretch{1.2}

\begin{table}[ht!]
	\caption{Average measures (with standard deviations) over 100 replications for Model 3. \textbf{Bold} letters indicate the best results. }
	\vspace{0.5cm}
	\label{table:Summary03}
	\begin{center}
		\resizebox{16cm}{!}{
			\begin{tabular}{c|@{\extracolsep{0.5cm}}c@{\extracolsep{0.5cm}}c@{\extracolsep{0.5cm}}c@{\extracolsep{0.5cm}}c@{\extracolsep{0.5cm}}c@{\extracolsep{0.5cm}}c@{\extracolsep{0.5cm}}c}
				\hline\hline
				
					& $\widehat{\Omega}_\text{GLasso}$  &  $\widehat{\Omega}_\text{GEN}$ & $\widehat{\Omega}_\text{T-GEN}$ & $\widehat{\Omega}_\text{GRidge}$
				& $\widehat{\Omega}_\text{T-GRidge}$  & $\widehat{\Omega}_\text{SCAD}$  & $\widehat{\Omega}_\text{EAGL}$   \\
				\hline
				KLL		& 37.53	(0.736)&	40.05	(0.420)&	37.814	(0.401)&	51.628	(0.608)&	56.159	(0.231)&	\textbf{34.445	(0.673)}&	37.047	(0.778) \\\hline
				
				RKLL	&86.52	(5.332)&	95.22	(1.214)&	92.807	(1.594)&	155.85	(2.930)&	144.915	(2.111)&	72.145	(3.845)&	\textbf{61.52	(0.917)} \\\hline
				
				$\ell_2$& 14.87	(0.214)&	15.10	(0.028)&	15.218	(0.044)&	16.479	(0.043)&	16.312	(0.033)&	14.396	(0.166) &	\textbf{13.93	(0.047)} \\\hline
				
				$\ell_{\text{sp}}$&3.249	(0.033)&	3.280	(0.008)&	3.299	(0.008)&	3.477	(0.007)&	3.448	(0.005)&	3.180	(0.026)&	\textbf{3.096	(0.011)} \\\hline
				
				$\ell_1$& 3.849	(0.078)&	3.942	(0.051)&	3.673	(0.043)&	3.602	(0.014)&	\textbf{3.525	(0.008)} &	3.758	(0.064)&	3.601	(0.057)	 \\\hline
				
				RTE		& 0.495	(0.018)&	0.514	(0.003)&	0.515	(0.005)&	0.603	(0.005)&	0.523	(0.006)&	0.447	(0.015) &	\textbf{0.359	(0.005)}	 \\\hline
				
				uMCC	& 0.538	(0.005)&	0.532	(0.004)&	0.550	(0.004)&	\textbf{0.599	(0.001)}&	0.594	(0.001)&	0.543	(0.004)&	0.550	(0.004)
				
				\\\hline
				
				BA		& 0.527	(0.003)&	0.527	(0.003)&	0.528	(0.002)&	0.526	(0.001)&	0.524	(0.001)&	0.528	(0.002)&	\textbf{0.528	(0.002)}
				\\\hline

				\hline
				\hline
			\end{tabular}
		}
	\end{center}
\end{table}

\renewcommand\arraystretch{1.2}

\begin{table}[ht!]
	\caption{Average measures (with standard deviations) over 100 replications for Model 4. \textbf{Bold} letters indicate the best results. }
	\vspace{0.5cm}
	\label{table:Summary04}
	\begin{center}
		\resizebox{16cm}{!}{
			\begin{tabular}{c|@{\extracolsep{0.5cm}}c@{\extracolsep{0.5cm}}c@{\extracolsep{0.5cm}}c@{\extracolsep{0.5cm}}c@{\extracolsep{0.5cm}}c@{\extracolsep{0.5cm}}c@{\extracolsep{0.5cm}}c}
				\hline\hline
				
					& $\widehat{\Omega}_\text{GLasso}$  &  $\widehat{\Omega}_\text{GEN}$ & $\widehat{\Omega}_\text{T-GEN}$ & $\widehat{\Omega}_\text{GRidge}$
				& $\widehat{\Omega}_\text{T-GRidge}$  & $\widehat{\Omega}_\text{SCAD}$  & $\widehat{\Omega}_\text{EAGL}$   \\
				\hline
				KLL		& 92.70	(1.535) &	95.88	(0.995) &	95.19	(1.002) &	135.1	(2.476) &	135.4	(1.873) &	\textbf{91.84	(1.019)} &	107.9	(1.891)\\\hline
				
				RKLL	&259.6	(22.24) &	267.8	(1.860) &	261.9	(1.884) &	289.5	(8.626) &	357.9	(2.342) &	264.4	(10.01) &	\textbf{155.8	(4.080)}\\\hline
				
				$\ell_2$& 17.26	(0.205) &	17.41	(0.013) &	17.36	(0.014) &	17.78	(0.054) &	18.23	(0.011) &	17.28	(0.107) &	\textbf{15.94	(0.077)}\\\hline
				
				$\ell_{\text{sp}}$&5.291	(0.024) &	5.291	(0.008) &	5.287	(0.008) &	5.236	(0.008) &	5.292	(0.005) &	5.303	(0.014) &	\textbf{5.150	(0.017)}\\\hline
				
				$\ell_1$&10.11	(0.061) &	10.11	(0.056) &	10.10	(0.057) &	10.01	(0.045) &	10.10	(0.039) &	10.11	(0.056) &	\textbf{9.913	(0.080)} \\\hline
				
				RTE		& 0.654	(0.017) &	0.678	(0.002) &	0.672	(0.002) &	0.713	(0.005) &	0.752	(0.001) &	0.649	(0.010) &	\textbf{0.481	(0.006)} \\\hline
				
				uMCC	& 0.597	(0.003) &	0.597	(0.003) &	0.598	(0.003) &	0.591	(0.002) &	0.594	(0.002) &	0.598	(0.003) &	\textbf{0.598	(0.003)}
				
				\\\hline
				
				BA		& 0.580	(0.003) &	0.583	(0.003) &	\textbf{0.583	(0.003)} &	0.532	(0.001) &	0.536	(0.002) &	0.578	(0.002) &	0.578	(0.003)
				\\\hline

				\hline
				\hline
			\end{tabular}
		}
	\end{center}
\end{table}

\newpage 

\renewcommand\arraystretch{1.2}

\begin{table}[ht!]
	\caption{Average measures (with standard deviations) over 100 replications for Model 5. \textbf{Bold} letters indicate the best results. }
	\vspace{0.5cm}
	\label{table:Summary05}
	\begin{center}
		\resizebox{16cm}{!}{
			\begin{tabular}{c|@{\extracolsep{0.5cm}}c@{\extracolsep{0.5cm}}c@{\extracolsep{0.5cm}}c@{\extracolsep{0.5cm}}c@{\extracolsep{0.5cm}}c@{\extracolsep{0.5cm}}c@{\extracolsep{0.5cm}}c}
				\hline\hline

				& $\widehat{\Omega}_\text{GLasso}$  &  $\widehat{\Omega}_\text{GEN}$ & $\widehat{\Omega}_\text{T-GEN}$ & $\widehat{\Omega}_\text{GRidge}$
				& $\widehat{\Omega}_\text{T-GRidge}$  & $\widehat{\Omega}_\text{SCAD}$  & $\widehat{\Omega}_\text{EAGL}$   \\
				\hline
				KLL		& 45.39	( 1.315) &	45.18	(1.020) &	\textbf{45.02	(0.835)} &	154.1	(2.849) &	178.4	(2.275) &	46.57	(1.179) &	46.80	(1.465)\\\hline
				
				RKLL	&61.58	( 4.690) &	64.58	(2.978) &	62.98	(0.962) &	132.5	(5.567) &	76.36	(1.454) &	55.74	(2.360) &	\textbf{46.15	(1.635)}\\\hline
				
				$\ell_2$& 8.147	( 0.310) &	8.287	(0.199) &	8.269	(0.039) &	10.87	(0.145) &	9.400	(0.057) &	7.946	(0.137) &	\textbf{7.552	(0.159)}\\\hline
				
				$\ell_{\text{sp}}$	&2.474	( 0.062) &	2.512	(0.039) &	2.510	(0.017) &	3.028	(0.016) &	2.868	(0.009) &	2.439	(0.033) &	\textbf{2.381	(0.046)}\\\hline
				
				$\ell_1$	&4.122	( 0.162) &	4.197	(0.157) &	4.156	(0.155) &	4.226	(0.034) &	4.074	(0.036) &	4.020	(0.158) &	\textbf{3.998	(0.162)} \\\hline
				
				RTE		& 0.334	( 0.034) &	0.355	(0.022) &	0.352	(0.004) &	0.551	(0.016) &	0.393	(0.007) &	0.298	(0.016) &	\textbf{0.246	(0.020)}\\\hline
				
				uMCC	& 0.669	(0.006) &	0.660	(0.005) &	0.664	(0.003) &	0.611	(0.004) &	0.611	(0.004 )&	\textbf{0.680	(0.004)} &	0.677	(0.004) 
				
				\\\hline
				
				BA		& 0.726	(0.005) &	0.727	(0.005) &	\textbf{0.728	(0.005)} &	0.528	(0.002) &	0.528	(0.002 )&	0.715	(0.006) &	0.718	(0.008) 
				\\\hline

				\hline
				\hline
			\end{tabular}
		}
	\end{center}
\end{table}

\renewcommand\arraystretch{1.2}

\begin{table}[ht!]
	\caption{Average measures (with standard deviations) over 100 replications for Model 6. \textbf{Bold} letters indicate the best results. }
	\vspace{0.5cm}
	\label{table:Summary06}
	\begin{center}
		\resizebox{16cm}{!}{
			\begin{tabular}{c|@{\extracolsep{0.5cm}}c@{\extracolsep{0.5cm}}c@{\extracolsep{0.5cm}}c@{\extracolsep{0.5cm}}c@{\extracolsep{0.5cm}}c@{\extracolsep{0.5cm}}c@{\extracolsep{0.5cm}}c}
				\hline\hline

			 	& $\widehat{\Omega}_\text{GLasso}$  &  $\widehat{\Omega}_\text{GEN}$ & $\widehat{\Omega}_\text{T-GEN}$ & $\widehat{\Omega}_\text{GRidge}$
				& $\widehat{\Omega}_\text{T-GRidge}$  & $\widehat{\Omega}_\text{SCAD}$  & $\widehat{\Omega}_\text{EAGL}$   \\
				\hline
				KLL		& 11.43	(0.555) &	13.43	(0.532) &	\textbf{7.243	(0.392)} &	23.27	(1.194) &	12.91	(0.391) &	9.222	(0.390) &	8.613	(0.432)\\\hline
				
				RKLL	&15.25	(0.881) &	18.99	(0.827) &	8.365	(0.797) &	15.27	(2.103) &	8.871	(0.849) &	10.15	(0.559) &	\textbf{7.977	(0.237)}\\\hline
				
				$\ell_2$& 4.552	(0.155) &	4.817	(0.105) &	\textbf{3.497	(0.148)} &	4.496	(0.296) &	3.616	(0.168) &	3.919	(0.085) &	3.586	(0.048)\\\hline
				
				$\ell_{\text{sp}}$	&1.014	(0.047) &	1.012	(0.035) &	\textbf{0.771	(0.029)} &	1.254	(0.031) &	1.005	(0.025) &	0.976	(0.052) &	0.895	(0.041)\\\hline
				
				$\ell_1$&3.436	(0.198) &	3.680	(0.203) &	3.136	(0.200) &	5.768	(0.222) &	4.984	(0.156) &	3.207	(0.213) &	\textbf{3.135	(0.208)}\\\hline
				
				RTE		& 0.211	(0.023) &	0.238	(0.013) &	0.163	(0.016) &	0.240	(0.030) &	0.164	(0.018) &	0.092	(0.021) &	\textbf{0.034	(0.008)}\\\hline
				
				uMCC	& 0.708	(0.023) &	0.639	(0.008) &	0.763	(0.017) &	\textbf{0.802	(0.005)} &	0.782	(0.016) &	0.787	(0.015) &	0.776	(0.013)
				
				\\\hline
				
				BA		& 0.831	(0.014) &	\textbf{0.851	(0.010)} &	0.774	(0.014) &	0.686	(0.006) &	0.747	(0.005) &	0.789	(0.014) &	0.796	(0.014)
				\\\hline

				\hline
				\hline
			\end{tabular}
		}
	\end{center}
\end{table}

\newpage

\renewcommand\arraystretch{1.2}

\begin{table}[ht!]
	\caption{Average measures (with standard deviations) over 100 replications for Model 7. \textbf{Bold} letters indicate the best results. }
	\vspace{0.5cm}
	\label{table:Summary07}
	\begin{center}
		\resizebox{16cm}{!}{
			\begin{tabular}{c|@{\extracolsep{0.5cm}}c@{\extracolsep{0.5cm}}c@{\extracolsep{0.5cm}}c@{\extracolsep{0.5cm}}c@{\extracolsep{0.5cm}}c@{\extracolsep{0.5cm}}c@{\extracolsep{0.5cm}}c}
				\hline\hline

					& $\widehat{\Omega}_\text{GLasso}$  &  $\widehat{\Omega}_\text{GEN}$ & $\widehat{\Omega}_\text{T-GEN}$ & $\widehat{\Omega}_\text{GRidge}$
				& $\widehat{\Omega}_\text{T-GRidge}$  & $\widehat{\Omega}_\text{SCAD}$  & $\widehat{\Omega}_\text{EAGL}$   \\
				\hline
				KLL		& 13.14	(0.615) &	15.57	(0.741) &	11.52	(0.598) &	25.03	(1.210) &	21.08	(0.655) &	9.753	(0.817) &	\textbf{9.842	(0.786)}\\\hline
				
				RKLL	&16.84	(0.837) &	21.10	(1.057) &	13.12	(0.747) &	20.94	(0.554) &	15.15	(0.544) &	10.21	(1.045) &	\textbf{8.391	(0.456)}\\\hline
				
				$\ell_2$&4.299	(0.140) &	4.642	(0.167) &	3.952	(0.136) &	4.977	(0.053) &	4.336	(0.094) &	3.198	(0.196) &	\textbf{2.922	(0.084)}\\\hline
				
				$\ell_{\text{sp}}$&0.974	(0.040) &	0.987	(0.032) &	0.915	(0.040) &	1.135	(0.021) &	1.019	(0.016) &	0.860	(0.061) &	\textbf{0.846	(0.040)}\\\hline
				
				$\ell_1$	&3.447	(0.148) &	3.729	(0.147) &	3.224	(0.165) &	3.803	(0.113) &	3.600	(0.100) &	\textbf{3.089	(0.189)} &	3.114	(0.181)\\\hline
				
				RTE			& 0.230	(0.018) &	0.251	(0.019) &	0.223	(0.012) &	0.260	(0.001) &	0.219	(0.010) &	0.085	(0.031) &	\textbf{0.061	(0.007)}\\\hline
				
				uMCC		& 0.744	(0.019) &	0.664	(0.012) &	0.798	(0.032) &	0.892	(0.011) &	\textbf{0.954	(0.009)} &	0.804	(0.021) &	0.815	(0.007)
				
				\\\hline
				
				BA			&0.976	(0.006) &	0.945	(0.007) &	0.983	(0.005) &	0.815	(0.018) &	0.969	(0.009) &	0.984	(0.003) &	\textbf{0.986	(0.003)}
				\\\hline

				\hline
				\hline
			\end{tabular}
		}
	\end{center}
\end{table}

\newpage

\newpage

\bibliographystyle{apalike}
\bibliography{jcgs-template}

\newpage

\setcounter{page}{1}
\section*{Supplementary Materials}

\appendix
\section{Technical details}  
\label{AppendixA}

Entropy is a crucial scientific concept closely related to uncertainty and randomness. It originates from the field of information theory and is introduced by \cite{shannon1948mathematical}. It is borrowed in statistics for understanding the randomness in the data. Suppose $\textbf{X}$ follows a multivariate Gaussian distribution with mean $\mu$ and covariance matrix $\Sigma = \Omega^{-1}$, i.e., $\textbf{X}\sim N({\mu}, \Omega^{-1}) $. The entropy of a continuous random variable is defined as: 
\begin{equation}
	H(\textbf{X}) = -  \int p(\textbf{X}) \log p(\textbf{X})\,dx, 
\end{equation}
where $p(\textbf{X})$ probability density function and is defined for the Gaussian distribution as:
\begin{equation}
	p(\textbf{X}) = (2\pi)^{-p/2} \det(\Omega)^{1/2} \exp\left( -\dfrac{1}{2}(\textbf{X}-\mu)^T\Omega(\textbf{X}-\mu)\right)
\end{equation}
Therefore, we have:
\begin{eqnarray*}
	H(\textbf{X}) & = & -  \textbf{E}[log (p(\textbf{X}))] \\
	 & = & -\textbf{E}\left[\log\left[(2\pi)^{-p/2} \det(\Omega)^{1/2} \exp\left( -\dfrac{1}{2}(\textbf{X}-\mu)^T\Omega(\textbf{X}-\mu)\right)\right]\right] \\
	 & = & \dfrac{p}{2}\log(2\pi) - \dfrac{1}{2}\log\det(\Omega) + \dfrac{1}{2}\textbf{E}\left[(\textbf{X}-\mu)^T\Omega(\textbf{X}-\mu)\right]\\
	 & = & \dfrac{p}{2}(1+\log(2\pi)) - \dfrac{1}{2}\log\det(\Omega)\\
	 & = & \dfrac{p}{2}(1+\log(2\pi)) + \dfrac{1}{2}\log\det(\Omega^{-1})\\
	 & = & \dfrac{p}{2}(1+\log(2\pi)) +  \dfrac{1}{2}H_C(\textbf{X})\\
	 & = & C + \dfrac{1}{2}H_C(\textbf{X}),
\end{eqnarray*}
where $C$ is a constant. Here, we used the fact that $\textbf{E}\left[(\textbf{X}-\mu)^T\Omega(\textbf{X}-\mu)\right]=p$.

\setcounter{figure}{0}

\setcounter{table}{0}

\newpage

\section{Simulation models}
\label{AppendixB}
	\vspace*{-0.5cm}
\begin{figure}[h!]
	\label{Fig3}
	\caption{Gaussian Graphical models corresponding to each precision matrix model.}

	\begin{center}
		\includegraphics[width=50mm]{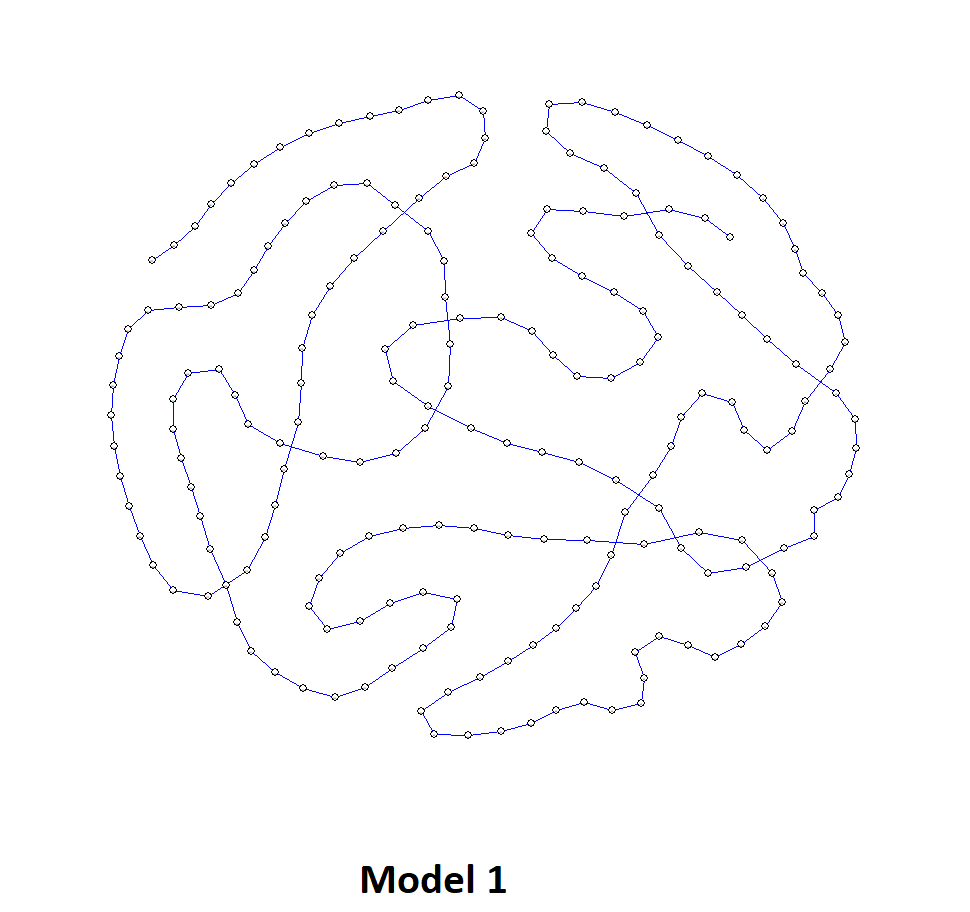}  \ \ \ \ \  
		\includegraphics[width=50mm]{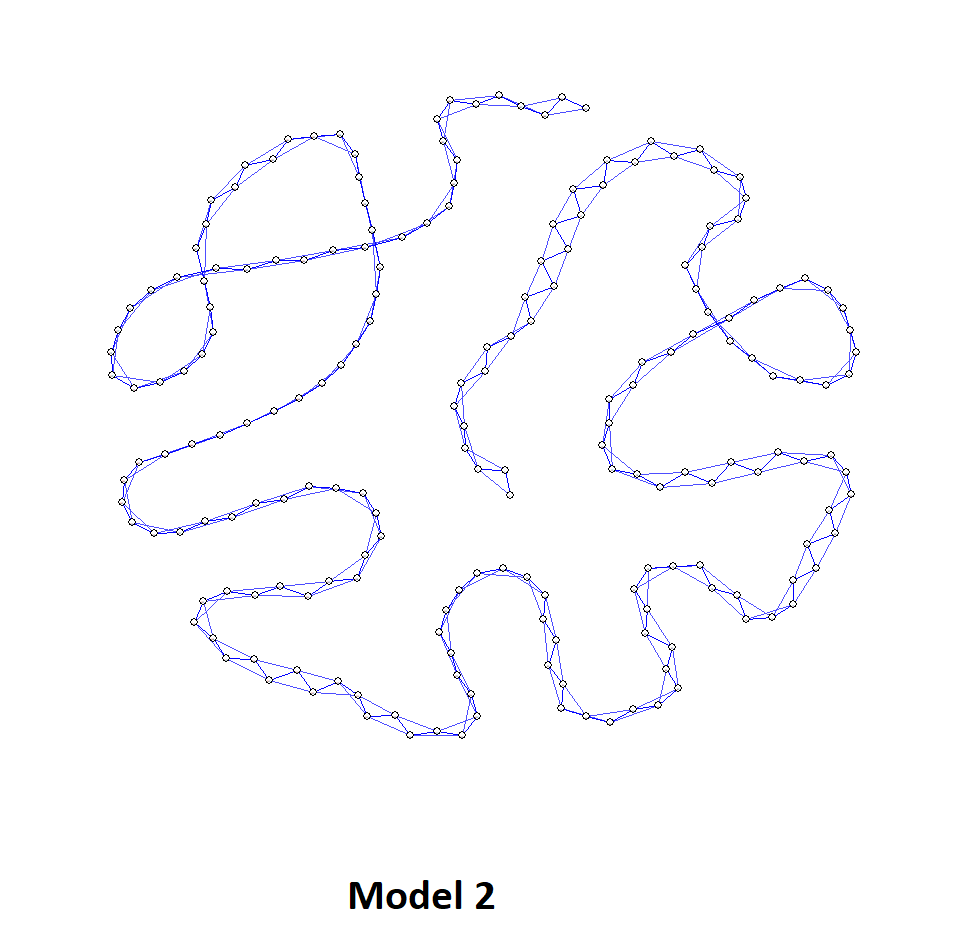}   \\  \bigskip
		\includegraphics[width=50mm]{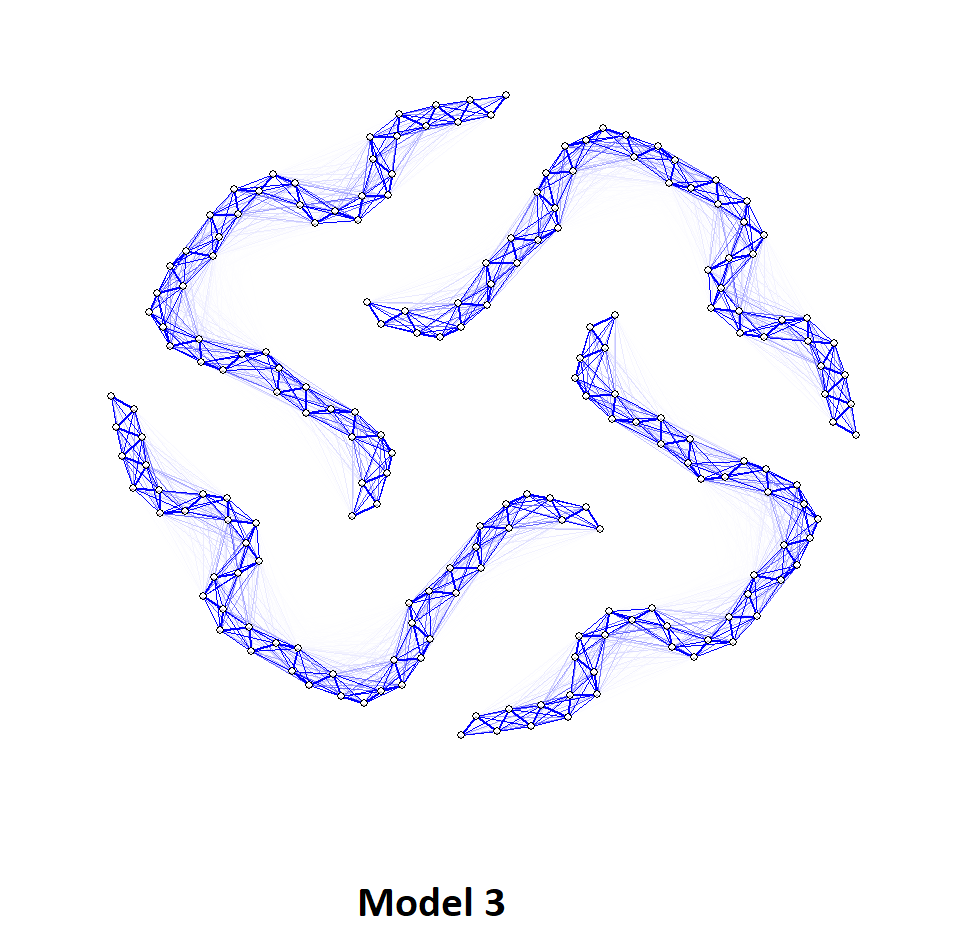}  \ \ \  \ \ 
		\includegraphics[width=50mm]{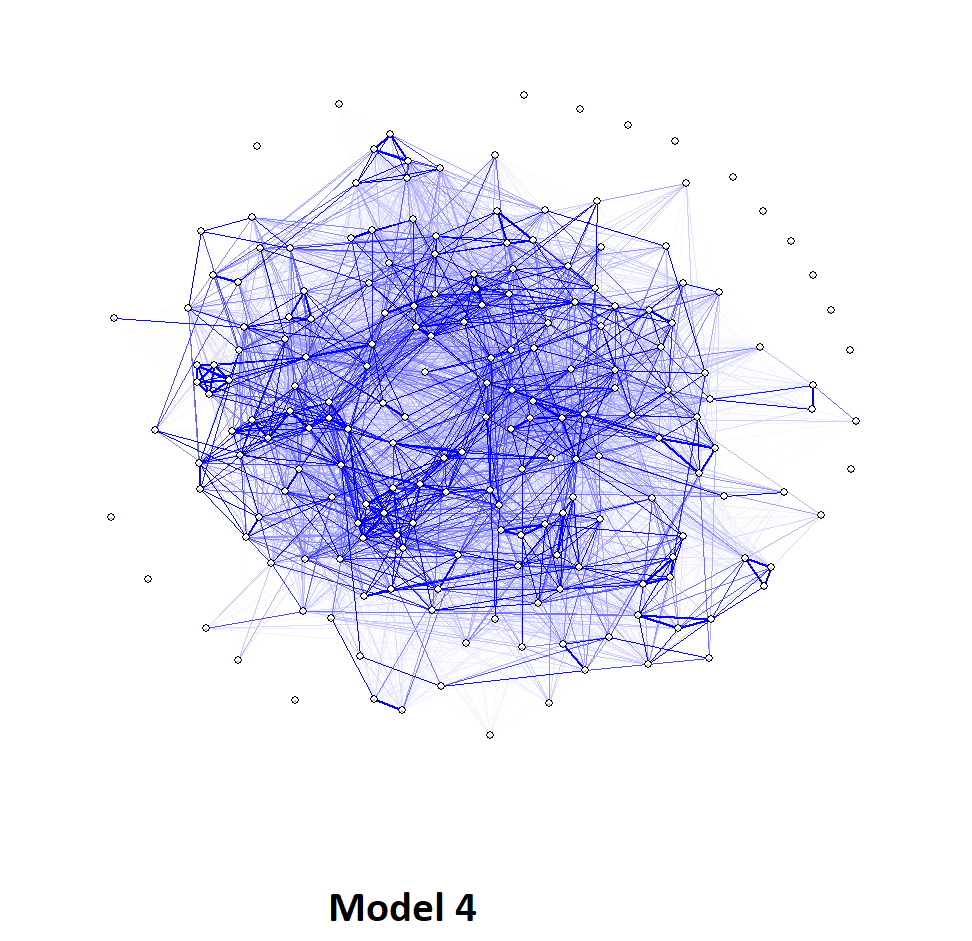}  \\ \bigskip
		\includegraphics[width=45mm]{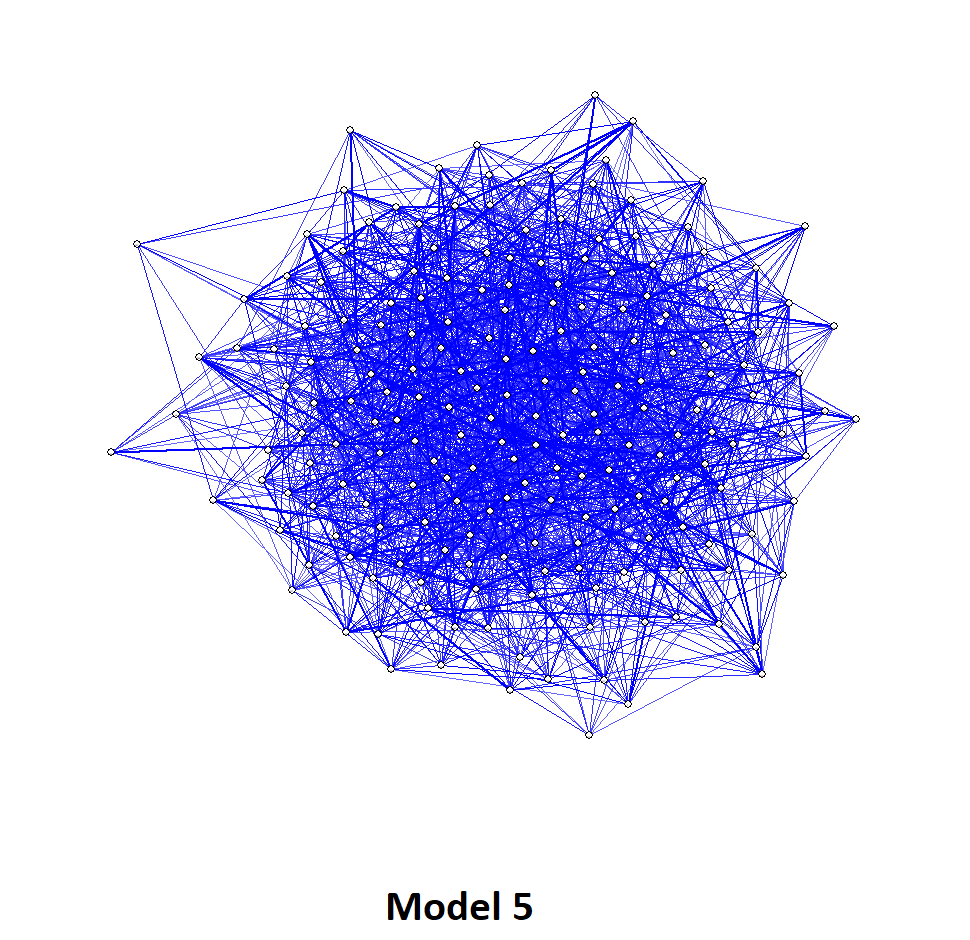}  \ \ \  \ \ 
		\includegraphics[width=45mm]{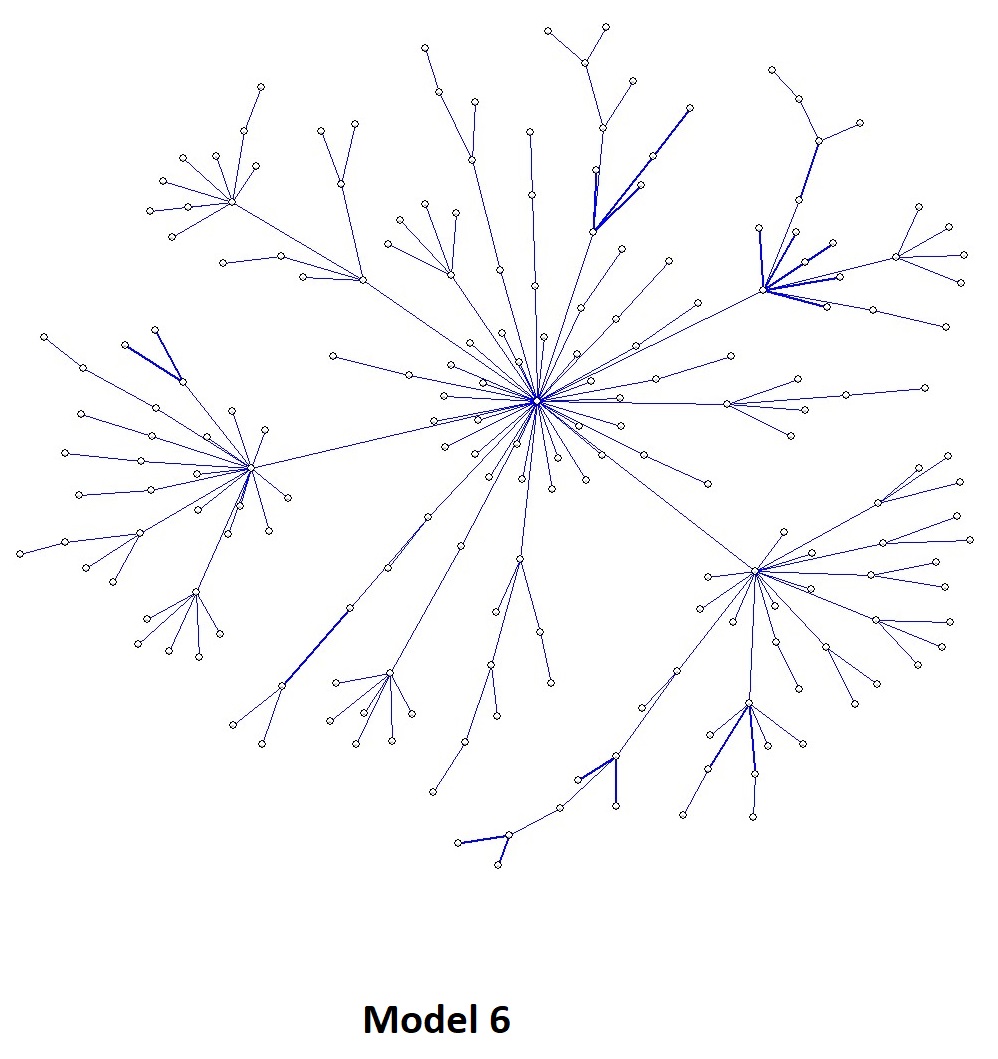}  \\ \ \bigskip
		\includegraphics[width=50mm]{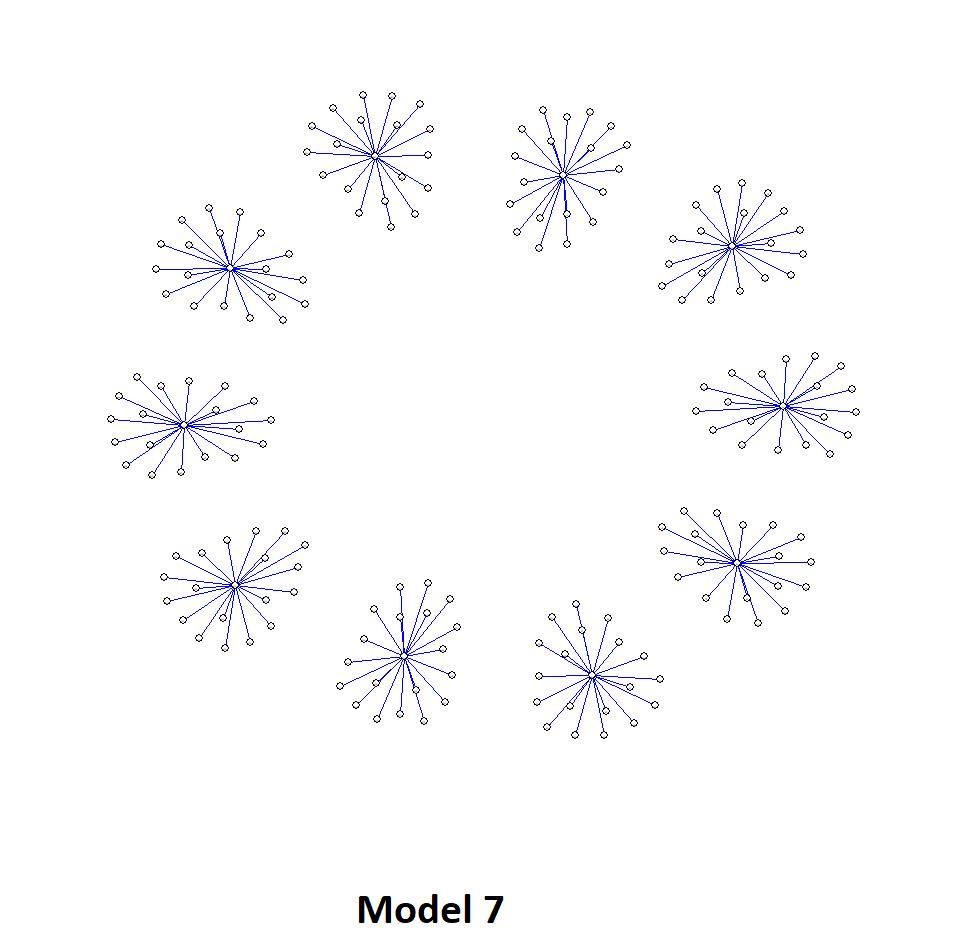}  \ \ \  \ \ 
		
	\end{center}
	
\end{figure}

\newpage

\section{Additional simulation results}
\label{AppendixC}

In this section, we present additional numerical results based on the same simulation settings described in Section \ref{section:4}, with the penalty parameter $\gamma$ now selected using the BIC criterion. For a comprehensive comparison, we provide the results for both 5-fold CV and BIC. The results indicate that, in the majority of cases, the precision matrices estimated using CV tend to have lower statistical losses compared to those estimated with BIC. This suggests that CV may offer a more effective approach for obtaining lower statistical losses in these scenarios.

When examining Gaussian Graphical Model (GGM) selection, the results are more mixed. For certain models, such as models 4 and 5, CV leads to higher values for both uMCC and BA, compared to BIC, suggesting better sparsity pattern selection performance. However, in other cases, such as models 2 and 6, CV results in higher BA but lower uMCC than BIC. Therefore, we do not observe a consistent advantage of BIC approach in terms of sparsity prediction measures.

\newpage

\renewcommand\arraystretch{1.2}

\begin{table}[ht!]
	\caption{Average measures (with standard deviations) over 100 replications for Model 1. \textbf{Bold} letters indicate the best results. }
	\vspace{0.5cm}
	\label{table:Summary1}
	\begin{center}
		\resizebox{16cm}{!}{
			\begin{tabular}{c|@{\extracolsep{0.5cm}}c|@{\extracolsep{0.5cm}}c@{\extracolsep{0.5cm}}c@{\extracolsep{0.5cm}}c@{\extracolsep{0.5cm}}c@{\extracolsep{0.5cm}}c@{\extracolsep{0.5cm}}c@{\extracolsep{0.5cm}}c}
				\hline\hline
				
				& 	& $\widehat{\Omega}_\text{GLasso}$  &  $\widehat{\Omega}_\text{GEN}$ & $\widehat{\Omega}_\text{T-GEN}$ & $\widehat{\Omega}_\text{GRidge}$
				& $\widehat{\Omega}_\text{T-GRidge}$  & $\widehat{\Omega}_\text{SCAD}$  & $\widehat{\Omega}_\text{EAGL}$   \\
				\hline
				KLL		& BIC	& {49.11	(4.926)}&   63.89	(2.843) & 53.67 (0.518) & 56.65	(1.556) & 59.18 (2.177) &  \textbf{36.45 (0.638)} &  36.71 (2.710) \\
				& CV	& 27.56 	(0.930) &	33.08	(0.627) & 30.86	(0.541) & 58.14	(1.399) & 64.03	(2.624) & \textbf{21.00 (0.591)} & 23.61	(0.839)  \\\hline
				
				RKLL	& BIC	& {89.66	(12.02)}&  132.2	(7.750) & 94.19	(2.016) & 49.01	(0.720) & 49.70 (0.799) & 58.29 (0.954) & \textbf{34.87	(2.007)} \\
				& CV	&  38.30	(2.767) &  49.72	(1.681) & 45.20 (0.918) & 75.97	(0.832) & 87.32	(5.451) & 25.49	(0.745) & \textbf{22.77	(0.845)} \\\hline
				
				$\ell_2$& BIC	& {10.48	(0.369)}&  {11.54	(0.163)}& 10.67	(0.060)		&	8.404	(0.031) & 8.427	(0.038) & 9.296	(0.042) & \textbf{7.517	(0.164)} \\
				& CV	& 7.493	(0.401) &	8.365	(0.191) & 	8.193	(0.047)  	&	9.994	(0.019) & 10.40	(0.201) & 6.318	(0.078) & \textbf{6.107	(0.153)} \\\hline
				
				$\ell_{\text{sp}}$&BIC	& 1.303	(0.034) &  1.407	(0.015) & 1.333	(0.013) & 1.086	(0.018) & 1.089	(0.015) & 1.190	(0.013)	& \textbf{1.005	(0.018)} \\
				& CV	& 1.041	(0.039) & 1.126	 (0.019)	& 1.105	(0.013) & 1.253	(0.010) & 1.308	(0.025) & 0.927	(0.023) &	\textbf{0.891	(0.023)} \\\hline
				
				$\ell_1$&BIC	& 1.472	(0.028) & 1.529	(0.021) & 	1.430	(0.018)	& 1.223	(0.039) &	\textbf{1.216	(0.032)} &	1.416	(0.039) &	1.219	(0.043)\\
				& CV	& 1.654	(0.084) & 1.775	(0.066) &	1.688	(0.053) & \textbf{1.356	(0.026)} &	1.395	(0.029) &	1.467	(0.068) &	1.403	(0.068) \\\hline
				
				RTE		& BIC	& 0.561	(0.022) &	0.624	(0.010)	& 0.560	(0.006) &	0.354	(0.001) &	0.346	(0.004) &	0.489	(0.003) & 	\textbf{0.328	(0.007)} \\
				& CV	& 0.369	(0.030) &	0.419	(0.015) & 0.410	(0.003) &	0.482	(0.001) &	0.495	(0.015) &	0.297	(0.005) &	\textbf{0.256	(0.010)} \\\hline
				
				uMCC	&BIC	& 0.818	(0.031) &	0.835	(0.017) &	0.894	(0.008) & \textbf{0.983	(0.005)} &	0.976	(0.009) &	0.787	(0.006) &	0.842	(0.019) \\
				& CV	& 0.653	(0.011) &	0.634	(0.002) &	0.648	(0.002) & \textbf{0.979	(0.007)} &	0.955	(0.018) &	0.681	(0.003) &	0.704	(0.004) \\\hline
				
				BA		&BIC	& 0.988	(0.003) &	0.990	(0.002) &	\textbf{0.994	(0.001)} &	0.989	(0.005) &	0.991	(0.006) &	0.985	(0.001) &	0.991	(0.002)\\
				& CV	& 0.936	(0.009) &	0.919	(0.002) &	0.933	(0.002) & 0.990	(0.006) & \textbf{0.990	(0.008)} & 0.955	(0.002) & 0.965	(0.001) 	 \\\hline

				\hline
				\hline
			\end{tabular}
		}
	\end{center}
\end{table}

\newpage

\renewcommand\arraystretch{1.2}

\begin{table}[ht!]
	\caption{Average measures (with standard deviations) over 100 replications for Model 2. \textbf{Bold} letters indicate the best results. }
	\vspace{0.5cm}
	\label{table:Summary2}
	\begin{center}
		\resizebox{16cm}{!}{
			\begin{tabular}{c|@{\extracolsep{0.5cm}}c|@{\extracolsep{0.5cm}}c@{\extracolsep{0.5cm}}c@{\extracolsep{0.5cm}}c@{\extracolsep{0.5cm}}c@{\extracolsep{0.5cm}}c@{\extracolsep{0.5cm}}c@{\extracolsep{0.5cm}}c}
				\hline\hline
				
				& 	& $\widehat{\Omega}_\text{GLasso}$  &  $\widehat{\Omega}_\text{GEN}$ & $\widehat{\Omega}_\text{T-GEN}$ & $\widehat{\Omega}_\text{GRidge}$
				& $\widehat{\Omega}_\text{T-GRidge}$  & $\widehat{\Omega}_\text{SCAD}$  & $\widehat{\Omega}_\text{EAGL}$   \\
				\hline
				KLL		& BIC	& 75.61	(4.260) &	84.73	(2.809) &	70.54	(1.776)	& 69.66	(1.576) & 71.54	(1.886) & 68.91	(1.495) & \textbf{66.71	(1.902)} \\
				& CV	&45.05	(1.137) &	47.27	(0.629) &	46.80	(0.706) &	69.07	(1.931) &	77.36	(0.849) &	\textbf{42.08	(1.267)} &	46.07	(1.135)  \\\hline
				
				RKLL	& BIC	& 203.5	(17.52)	& 250.6	(13.05) &	159.0	(3.829) &	135.5	(2.541) &	96.57	(1.708) &	170.8	(5.746) &	\textbf{90.29	(0.762)} \\
				& CV	& 77.94	(6.640) &	84.77	(0.981) &	85.06	(3.230) &	106.7	(3.601) &	162.0	(2.687) &	63.59	(5.960) &\textbf{58.75	(1.691)} \\\hline
				
				$\ell_2$& BIC	& 14.79	(0.185) &	15.24 	(0.113) &	14.25 	(0.056) &	13.83	(0.043) &	12.85	(0.048) &	14.41	(0.069) &	\textbf{12.68	(0.021)}\\
				& CV	& 11.92	(0.393) &	12.17	(0.029) &	12.29	(0.180) &	13.15	(0.085) &	14.31	(0.040) &	11.27	(0.374) &	\textbf{11.21	(0.147)}\\\hline
				
				$\ell_{\text{sp}}$&BIC	& 2.313	(0.021) &	2.361	(0.013) &	2.253	(0.009) &	2.181 (0.010) &	2.059	(0.012) &	2.271	(0.009) &	\textbf{2.059 (0.009)} \\
				& CV	&1.954	(0.054) &	1.984	(0.010) &	2.002	(0.025) &	2.096	(0.014) &	2.247	(0.006) &	1.875	(0.054) &	\textbf{1.867	(0.023)} \\\hline
				
				$\ell_1$&BIC	& 2.452	(0.020) &	2.462	(0.015)	& 2.382	(0.021) &	2.310	(0.022) &	\textbf{2.212	(0.027)} &	2.443	(0.024) &	2.307	(0.036) \\
				
				& CV	& 2.683	(0.067) &	2.773	(0.043) &	2.695	(0.050) &	\textbf{2.250	(0.027)} &	2.325	(0.014) &	2.602	(0.058) &	2.547	(0.060) \\\hline
				
				RTE		& BIC	& 0.646	(0.011) &	0.682	(0.007) &	0.585	(0.006) &	0.560	(0.001) &	0.448	(0.004) &	0.616	(0.005) &	\textbf{0.419	(0.004)}\\
				& CV	&0.467	(0.028) &	0.484	(0.002) &	0.489	(0.013) &	0.493	(0.007) &	0.573	(0.005) &	0.412	(0.026) &	\textbf{0.366	(0.010)}\\\hline
				
				uMCC	&BIC	& 0.822	(0.008) &	0.827	(0.007) &	0.829	(0.006) &	\textbf{0.871	(0.009)} &	0.866	(0.008) &	0.817	(0.005) &	0.817	(0.005)
				\\
				& CV	&0.671	(0.011) &	0.651	(0.002) &	0.668	(0.006) &	\textbf{0.873	(0.008)} &	0.827	(0.007) &	0.689	(0.009) &	0.703	(0.007) \\\hline
				
				BA		&BIC	& 0.892	(0.025) &	0.873	(0.018) &	0.861	(0.018) &	0.830	(0.017) &	0.830	(0.015) &	0.913	(0.011) & 	\textbf{0.914}	(0.008)
				\\
				& CV	& 0.919	(0.010) &	0.900	(0.002) &	0.917	(0.005) &	0.829	(0.016) &	0.795	(0.014) &	0.933	(0.005) &	\textbf{0.940	(0.004)} \\\hline

				\hline
				\hline
			\end{tabular}
		}
	\end{center}
\end{table}

\newpage

\renewcommand\arraystretch{1.2}

\begin{table}[ht!]
	\caption{Average measures (with standard deviations) over 100 replications for Model 3. \textbf{Bold} letters indicate the best results. }
	\vspace{0.5cm}
	\label{table:Summary3}
	\begin{center}
		\resizebox{16cm}{!}{
			\begin{tabular}{c|@{\extracolsep{0.5cm}}c|@{\extracolsep{0.5cm}}c@{\extracolsep{0.5cm}}c@{\extracolsep{0.5cm}}c@{\extracolsep{0.5cm}}c@{\extracolsep{0.5cm}}c@{\extracolsep{0.5cm}}c@{\extracolsep{0.5cm}}c}
				\hline\hline
				
				& 	& $\widehat{\Omega}_\text{GLasso}$  &  $\widehat{\Omega}_\text{GEN}$ & $\widehat{\Omega}_\text{T-GEN}$ & $\widehat{\Omega}_\text{GRidge}$
				& $\widehat{\Omega}_\text{T-GRidge}$  & $\widehat{\Omega}_\text{SCAD}$  & $\widehat{\Omega}_\text{EAGL}$   \\
				\hline
				KLL		& BIC	& 52.57	(0.767)	& 58.64 (0.327) &	43.46 	(0.432) &	\textbf{41.85	(0.878)} &	45.48 	(0.942)  &	44.87 	(3.644) &	46.53 (2.486) \\
				& CV	& 37.53	(0.736)&	40.05	(0.420)&	37.814	(0.401)&	51.628	(0.608)&	56.159	(0.231)&	\textbf{34.445	(0.673)}&	37.047	(0.778) \\\hline
				
				RKLL	& BIC	& 158.2	(3.365) &	189.0	(1.541) &	111.0 (1.337) &	86.20	(1.480) &	80.51	(1.174) &	122.1	(15.86) &	\textbf{77.41	(3.666)} \\
				& CV	&86.52	(5.332)&	95.22	(1.214)&	92.807	(1.594)&	155.85	(2.930)&	144.915	(2.111)&	72.145	(3.845)&	\textbf{61.52	(0.917)} \\\hline
				
				$\ell_2$& BIC	&  16.48  (0.050) &	16.85 	(0.016) &	15.68	(0.029) &	15.01 	(0.039) &	14.76	(0.038) &	15.87	(0.300) &	\textbf{14.61	(0.138)}\\
				& CV	& 14.87	(0.214)&	15.10	(0.028)&	15.218	(0.044)&	16.479	(0.043)&	16.312	(0.033)&	14.396	(0.166) &	\textbf{13.93	(0.047)} \\\hline
				
				$\ell_{\text{sp}}$&BIC	& 3.480	(0.008) &	3.529	(0.004) &	3.368	(0.008) &	3.262	(0.008) &	3.216	(0.008) &	3.395	(0.043) &	\textbf{3.192	(0.019)} \\
				& CV	&3.249	(0.033)&	3.280	(0.008)&	3.299	(0.008)&	3.477	(0.007)&	3.448	(0.005)&	3.180	(0.026)&	\textbf{3.096	(0.011)} \\\hline
				
				$\ell_1$&BIC	& 3.630	(0.019) &	3.658	(0.016)	& 3.526	(0.015) &	3.441	(0.022) &	\textbf{3.389	(0.021)} &	3.608	(0.029) &	3.417	(0.032)\\
				
				& CV	& 3.849	(0.078)&	3.942	(0.051)&	3.673	(0.043)&	3.602	(0.014)&	\textbf{3.525	(0.008)} &	3.758	(0.064)&	3.601	(0.057)	 \\\hline
				
				RTE		& BIC	& 0.607	(0.004) &	0.641	(0.002) &	0.523	(0.006) &	0.441	(0.001) &	0.391	(0.004) &	0.557	(0.021) &	\textbf{0.371	(0.005)}\\
				& CV	& 0.495	(0.018)&	0.514	(0.003)&	0.515	(0.005)&	0.603	(0.005)&	0.523	(0.006)&	0.447	(0.015) &	\textbf{0.359	(0.005)}	 \\\hline
				
				uMCC	&BIC	& 0.590	(0.003)	& 0.590	(0.002) &	0.590	(0.002)  &	\textbf{0.599	(0.001)} &	0.598	(0.001) &	0.579	(0.009) &	0.587	(0.007)				
				\\
				& CV	& 0.538	(0.005)&	0.532	(0.004)&	0.550	(0.004)&	\textbf{0.599	(0.001)}&	0.594	(0.001)&	0.543	(0.004)&	0.550	(0.004)
				
				\\\hline
				
				BA		&BIC	& 0.527	(0.001) &	0.528	(0.001) &	0.528	(0.001) &	0.526	(0.001) &	0.526	(0.001) &	\textbf{0.528	(0.001)} &	0.528	(0.001)				
				\\
				& CV	& 0.527	(0.003)&	0.527	(0.003)&	0.528	(0.002)&	0.526	(0.001)&	0.524	(0.001)&	0.528	(0.002)&	\textbf{0.528	(0.002)}
				\\\hline

				\hline
				\hline
			\end{tabular}
		}
	\end{center}
\end{table}

\newpage

\renewcommand\arraystretch{1.2}

\begin{table}[ht!]
	\caption{Average measures (with standard deviations) over 100 replications for Model 4. \textbf{Bold} letters indicate the best results. }
	\vspace{0.5cm}
	\label{table:Summary4}
	\begin{center}
		\resizebox{16cm}{!}{
			\begin{tabular}{c|@{\extracolsep{0.5cm}}c|@{\extracolsep{0.5cm}}c@{\extracolsep{0.5cm}}c@{\extracolsep{0.5cm}}c@{\extracolsep{0.5cm}}c@{\extracolsep{0.5cm}}c@{\extracolsep{0.5cm}}c@{\extracolsep{0.5cm}}c}
				\hline\hline
				
				& 	& $\widehat{\Omega}_\text{GLasso}$  &  $\widehat{\Omega}_\text{GEN}$ & $\widehat{\Omega}_\text{T-GEN}$ & $\widehat{\Omega}_\text{GRidge}$
				& $\widehat{\Omega}_\text{T-GRidge}$  & $\widehat{\Omega}_\text{SCAD}$  & $\widehat{\Omega}_\text{EAGL}$   \\
				\hline
				KLL		& BIC	& 132.9	(5.040) &	141.4	(4.639) &	145.6	(3.649) &	137.1	(2.930) &	138.6	(2.934) &	133.1	(4.732) &	\textbf{132.8	(3.981)}\\
				& CV	& 92.70	(1.535) &	95.88	(0.995) &	95.19	(1.002) &	135.1	(2.476) &	135.4	(1.873) &	\textbf{91.84	(1.019)} &	107.9	(1.891)\\\hline
				
				RKLL	& BIC	& 716.2	(57.46) &	783.4	(55.54) &	791.5	(35.23) &	257.8	(2.257) &	248.1	(2.498) &	718.6	(53.93) &	\textbf{202.0	(3.505)}\\
				& CV	&259.6	(22.24) &	267.8	(1.860) &	261.9	(1.884) &	289.5	(8.626) &	357.9	(2.342) &	264.4	(10.01) &	\textbf{155.8	(4.080)}\\\hline
				
				$\ell_2$& BIC	&  19.10	(0.107) &	19.27	(0.088) &	19.31	(0.059) &	17.51	(0.012) &	17.41	(0.013) &	19.10	(0.101) &	\textbf{16.62	(0.038)}\\
				& CV	& 17.26	(0.205) &	17.41	(0.013) &	17.36	(0.014) &	17.78	(0.054) &	18.23	(0.011) &	17.28	(0.107) &	\textbf{15.94	(0.077)}\\\hline
				
				$\ell_{\text{sp}}$&BIC	& 5.478	(0.010) &	5.484	(0.009) &	5.473	(0.005) &	5.202	(0.007) &	\textbf{5.190	(0.007)} &	5.478	(0.009) &	5.248	(0.010) \\
				& CV	&5.291	(0.024) &	5.291	(0.008) &	5.287	(0.008) &	5.236	(0.008) &	5.292	(0.005) &	5.303	(0.014) &	\textbf{5.150	(0.017)}\\\hline
				
				$\ell_1$&BIC	& 10.35	(0.028) &	10.37	(0.026) &	10.37	(0.024) &	9.961	(0.049) &	\textbf{9.944	(0.049)} &	10.36	(0.028) &	9.994	(0.058)\\
				
				& CV	&10.11	(0.061) &	10.11	(0.056) &	10.10	(0.057) &	10.01	(0.045) &	10.10	(0.039) &	10.11	(0.056) &	\textbf{9.913	(0.080)} \\\hline
				
				RTE		& BIC	& 0.811	(0.010) &	0.831	(0.008) &	0.836	(0.005) &	0.688	(0.001) &	0.678	(0.001) &	0.812	(0.009) &	\textbf{0.508	(0.003)}\\
				& CV	& 0.654	(0.017) &	0.678	(0.002) &	0.672	(0.002) &	0.713	(0.005) &	0.752	(0.001) &	0.649	(0.010) &	\textbf{0.481	(0.006)} \\\hline
				
				uMCC	&BIC	& 0.586	(0.003) &	0.586	(0.003) &	0.581	(0.003) &	0.590	(0.002) &	0.591	(0.002) &	0.586	(0.003) &	\textbf{0.592	(0.003) }		
				\\
				& CV	& 0.597	(0.003) &	0.597	(0.003) &	0.598	(0.003) &	0.591	(0.002) &	0.594	(0.002) &	0.598	(0.003) &	\textbf{0.598	(0.003)}
				
				\\\hline
				
				BA		&BIC	& 0.545	(0.004) &	0.544	(0.003) &	0.537	(0.003) &	0.532	(0.001) &	0.532	(0.001) &	0.544	(0.003) &	\textbf{0.555	(0.003)}
				
				\\
				& CV	& 0.580	(0.003) &	0.583	(0.003) &	\textbf{0.583	(0.003)} &	0.532	(0.001) &	0.536	(0.002) &	0.578	(0.002) &	0.578	(0.003)
				\\\hline

				\hline
				\hline
			\end{tabular}
		}
	\end{center}
\end{table}

\newpage 

\renewcommand\arraystretch{1.2}

\begin{table}[ht!]
	\caption{Average measures (with standard deviations) over 100 replications for Model 5. \textbf{Bold} letters indicate the best results. }
	\vspace{0.5cm}
	\label{table:Summary5}
	\begin{center}
		\resizebox{16cm}{!}{
			\begin{tabular}{c|@{\extracolsep{0.5cm}}c|@{\extracolsep{0.5cm}}c@{\extracolsep{0.5cm}}c@{\extracolsep{0.5cm}}c@{\extracolsep{0.5cm}}c@{\extracolsep{0.5cm}}c@{\extracolsep{0.5cm}}c@{\extracolsep{0.5cm}}c}
				\hline\hline

				& 	& $\widehat{\Omega}_\text{GLasso}$  &  $\widehat{\Omega}_\text{GEN}$ & $\widehat{\Omega}_\text{T-GEN}$ & $\widehat{\Omega}_\text{GRidge}$
				& $\widehat{\Omega}_\text{T-GRidge}$  & $\widehat{\Omega}_\text{SCAD}$  & $\widehat{\Omega}_\text{EAGL}$   \\
				\hline
				KLL		& BIC	& 76.20	(5.512) &	87.71	(6.936) &	106.05	(3.101) &	149.8	(1.073) &	139.7	(1.820) &	\textbf{57.99	(3.468)} &	71.23	(4.913) \\
				& CV	& 45.39	( 1.315) &	45.18	(1.020) &	\textbf{45.02	(0.835)} &	154.1	(2.849) &	178.4	(2.275) &	46.57	(1.179) &	46.80	(1.465)\\\hline
				
				RKLL	& BIC	& 152.5	(16.89) &	195.3	(21.95) &	195.4	(12.77) &	199.5	(0.672) &	173.1	(9.354) &	93.04	(9.484) &	\textbf{65.86	(1.833)} \\
				& CV	&61.58	( 4.690) &	64.58	(2.978) &	62.98	(0.962) &	132.5	(5.567) &	76.36	(1.454) &	55.74	(2.360) &	\textbf{46.15	(1.635)}\\\hline
				
				$\ell_2$& BIC	&  10.95	(0.273) &	11.58	(0.256) &	11.61	(0.140) &	11.82	(0.009) &	11.50	(0.121) &	9.610	(0.298) &	\textbf{8.766	(0.078)} \\
				& CV	& 8.147	( 0.310) &	8.287	(0.199) &	8.269	(0.039) &	10.87	(0.145) &	9.400	(0.057) &	7.946	(0.137) &	\textbf{7.552	(0.159)}\\\hline
				
				$\ell_{\text{sp}}$&BIC	&2.956	(0.041) &	3.038	(0.036) &	3.065	(0.016) &	3.117	(0.004) &	3.077	(0.011) &	2.754	(0.051) &	\textbf{2.704	(0.024) } \\
				& CV	&2.474	( 0.062) &	2.512	(0.039) &	2.510	(0.017) &	3.028	(0.016) &	2.868	(0.009) &	2.439	(0.033) &	\textbf{2.381	(0.046)}\\\hline

				$\ell_1$&BIC	& 4.305	(0.091) &	4.372	(0.083) &	4.407	(0.110) &	4.307	(0.031) &	4.246	(0.027) &	\textbf{4.155	(0.106)} &	4.214	(0.160) \\
				
				& CV	&4.122	( 0.162) &	4.197	(0.157) &	4.156	(0.155) &	4.226	(0.034) &	4.074	(0.036) &	4.020	(0.158) &	\textbf{3.998	(0.162)} \\\hline
				
				RTE		& BIC	& 0.583	(0.022) &	0.637	(0.020) &	0.630	(0.013) &	0.642	(0.001) &	0.615	(0.011) &	0.465	(0.027) &	\textbf{0.348	(0.006)} \\
				& CV	& 0.334	( 0.034) &	0.355	(0.022) &	0.352	(0.004) &	0.551	(0.016) &	0.393	(0.007) &	0.298	(0.016) &	\textbf{0.246	(0.020)}\\\hline
				
				uMCC	&BIC	& 0.659	(0.007) &	0.651	(0.008) &	0.615	(0.005) &	0.615	(0.005) &	0.639	(0.004) &	\textbf{0.674	(0.005)} &	0.647	(0.007) 		
				\\
				& CV	& 0.669	(0.006) &	0.660	(0.005) &	0.664	(0.003) &	0.611	(0.004) &	0.611	(0.004 )&	\textbf{0.680	(0.004)} &	0.677	(0.004) 
				
				\\\hline
				
				BA		&BIC	& 0.648	(0.012) &	0.641	(0.014) &	0.566	(0.006) &	0.530	(0.002) &	0.554	(0.005) &	\textbf{0.678	(0.010)} &	0.626	(0.011) 
				
				\\
				& CV	& 0.726	(0.005) &	0.727	(0.005) &	\textbf{0.728	(0.005)} &	0.528	(0.002) &	0.528	(0.002 )&	0.715	(0.006) &	0.718	(0.008) 
				\\\hline

				\hline
				\hline
			\end{tabular}
		}
	\end{center}
\end{table}

\newpage 

\renewcommand\arraystretch{1.2}

\begin{table}[ht!]
	\caption{Average measures (with standard deviations) over 100 replications for Model 6. \textbf{Bold} letters indicate the best results. }
	\vspace{0.5cm}
	\label{table:Summary6}
	\begin{center}
		\resizebox{16cm}{!}{
			\begin{tabular}{c|@{\extracolsep{0.5cm}}c|@{\extracolsep{0.5cm}}c@{\extracolsep{0.5cm}}c@{\extracolsep{0.5cm}}c@{\extracolsep{0.5cm}}c@{\extracolsep{0.5cm}}c@{\extracolsep{0.5cm}}c@{\extracolsep{0.5cm}}c}
				\hline\hline

				& 	& $\widehat{\Omega}_\text{GLasso}$  &  $\widehat{\Omega}_\text{GEN}$ & $\widehat{\Omega}_\text{T-GEN}$ & $\widehat{\Omega}_\text{GRidge}$
				& $\widehat{\Omega}_\text{T-GRidge}$  & $\widehat{\Omega}_\text{SCAD}$  & $\widehat{\Omega}_\text{EAGL}$   \\
				\hline
				KLL		& BIC	& 13.68	(0.505) &	21.59	(1.415) &	8.404	(0.553) &	22.68	(1.165) &	13.15	(0.609) &	7.783	(0.366) &	\textbf{7.279	(0.493)}\\
				& CV	& 11.43	(0.555) &	13.43	(0.532) &	\textbf{7.243	(0.392)} &	23.27	(1.194) &	12.91	(0.391) &	9.222	(0.390) &	8.613	(0.432)\\\hline
				
				RKLL	& BIC	& 19.04	(0.737) &	32.27	(2.449) &	9.360	(1.114) &	13.50	(0.141) &	7.694	(0.524) &	9.186	(0.433) &	\textbf{6.353	(0.184)} \\
				& CV	&15.25	(0.881) &	18.99	(0.827) &	8.365	(0.797) &	15.27	(2.103) &	8.871	(0.849) &	10.15	(0.559) &	\textbf{7.977	(0.237)}\\\hline
				
				$\ell_2$& BIC	&  4.880	(0.075) &	5.971	(0.170) &	3.613	(0.175) &	4.281	(0.020) &	3.351	(0.120) &	3.576	(0.072) &	\textbf{3.070	(0.031)}\\
				& CV	& 4.552	(0.155) &	4.817	(0.105) &	\textbf{3.497	(0.148)} &	4.496	(0.296) &	3.616	(0.168) &	3.919	(0.085) &	3.586	(0.048)\\\hline
				
				$\ell_{\text{sp}}$&BIC	&1.005	(0.020) &	1.087	(0.024) &	\textbf{0.806	(0.023)} &	1.240	(0.030) &	0.918	(0.020) &	0.884	(0.027) &	0.818	(0.028) \\
				& CV	&1.014	(0.047) &	1.012	(0.035) &	\textbf{0.771	(0.029)} &	1.254	(0.031) &	1.005	(0.025) &	0.976	(0.052) &	0.895	(0.041)\\\hline
				
				$\ell_1$&BIC	& 3.932	(0.170) &	4.387	(0.151) &	3.065	(0.162) &	5.764	(0.227) &	4.024	(0.111) &	3.286	(0.206) &	\textbf{2.931	(0.179)} \\
				
				& CV	&3.436	(0.198) &	3.680	(0.203) &	3.136	(0.200) &	5.768	(0.222) &	4.984	(0.156) &	3.207	(0.213) &	\textbf{3.135	(0.208)}\\\hline
				
				RTE		& BIC	& 0.278	(0.007) &	0.373	(0.014) &	0.169	(0.018) &	0.221	(0.001) &	0.133	(0.015) &	0.122	(0.011) &	\textbf{0.031	(0.007)} \\
				& CV	& 0.211	(0.023) &	0.238	(0.013) &	0.163	(0.016) &	0.240	(0.030) &	0.164	(0.018) &	0.092	(0.021) &	\textbf{0.034	(0.008)}\\\hline
				
				uMCC	&BIC	& 0.777	(0.007) &	0.771	(0.010) &	0.812	(0.008) &	0.801	(0.005) &	\textbf{0.840	(0.004)} &	0.777	(0.007) &	0.800	(0.009)
				\\
				& CV	& 0.708	(0.023) &	0.639	(0.008) &	0.763	(0.017) &	\textbf{0.802	(0.005)} &	0.782	(0.016) &	0.787	(0.015) &	0.776	(0.013)
				
				\\\hline
				
				BA		&BIC	& 0.765	(0.006) &	\textbf{0.768	(0.009)} &	0.736	(0.002) &	0.685	(0.006) &	0.735	(0.005) &	0.765	(0.006) &	0.739	(0.005)
				
				\\
				& CV	& 0.831	(0.014) &	\textbf{0.851	(0.010)} &	0.774	(0.014) &	0.686	(0.006) &	0.747	(0.005) &	0.789	(0.014) &	0.796	(0.014)
				\\\hline

		\hline
				\hline
			\end{tabular}
		}
	\end{center}
\end{table}

\newpage 

\renewcommand\arraystretch{1.2}

\begin{table}[ht!]
	\caption{Average measures (with standard deviations) over 100 replications for Model 7. \textbf{Bold} letters indicate the best results. }
	\vspace{0.5cm}
	\label{table:Summary7}
	\begin{center}
		\resizebox{16cm}{!}{
			\begin{tabular}{c|@{\extracolsep{0.5cm}}c|@{\extracolsep{0.5cm}}c@{\extracolsep{0.5cm}}c@{\extracolsep{0.5cm}}c@{\extracolsep{0.5cm}}c@{\extracolsep{0.5cm}}c@{\extracolsep{0.5cm}}c@{\extracolsep{0.5cm}}c}
				\hline\hline

				& 	& $\widehat{\Omega}_\text{GLasso}$  &  $\widehat{\Omega}_\text{GEN}$ & $\widehat{\Omega}_\text{T-GEN}$ & $\widehat{\Omega}_\text{GRidge}$
				& $\widehat{\Omega}_\text{T-GRidge}$  & $\widehat{\Omega}_\text{SCAD}$  & $\widehat{\Omega}_\text{EAGL}$   \\
				\hline
				KLL		& BIC	& 19.27	(1.496) &	27.51	(0.994) &	15.07	(0.910) &	24.848	(1.504) &	19.50	(1.157) &	11.89	(0.898) &	\textbf{13.058	(1.210)}\\
				& CV	& 13.14	(0.615) &	15.57	(0.741) &	11.52	(0.598) &	25.03	(1.210) &	21.08	(0.655) &	9.753	(0.817) &	\textbf{9.842	(0.786)}\\\hline
				
				RKLL	& BIC	& 26.58	(2.464) &	41.24	(1.744) &	16.66	(1.040) &	14.01	(0.496) &	11.19	(0.395) &	13.78	(1.054) &	\textbf{10.63	(0.740)} \\
				& CV	&16.84	(0.837) &	21.10	(1.057) &	13.12	(0.747) &	20.94	(0.554) &	15.15	(0.544) &	10.21	(1.045) &	\textbf{8.391	(0.456)}\\\hline
				
				$\ell_2$& BIC	& 5.406	(0.195) &	6.499	(0.094) &	4.470	(0.120) &	3.933	(0.074) &	3.519	(0.071) &	3.837	(0.147) &	\textbf{3.317	(0.112)}\\
				& CV	&4.299	(0.140) &	4.642	(0.167) &	3.952	(0.136) &	4.977	(0.053) &	4.336	(0.094) &	3.198	(0.196) &	\textbf{2.922	(0.084)}\\\hline
				
				$\ell_{\text{sp}}$&BIC	&1.154	(0.043) &	1.239	(0.027) &	1.087	(0.041) &	1.039	(0.023) &	\textbf{0.952	(0.027)} &	1.006	(0.043) &	0.987	(0.046)\\
				& CV	&0.974	(0.040) &	0.987	(0.032) &	0.915	(0.040) &	1.135	(0.021) &	1.019	(0.016) &	0.860	(0.061) &	\textbf{0.846	(0.040)}\\\hline
				
				$\ell_1$&BIC	& 3.648	(0.149) &	3.827	(0.111) &	3.517	(0.184) &	3.715	(0.124) &	3.383	(0.141) &	\textbf{3.330	(0.177)} &	3.407	(0.204) \\
				
				& CV	&3.447	(0.148) &	3.729	(0.147) &	3.224	(0.165) &	3.803	(0.113) &	3.600	(0.100) &	\textbf{3.089	(0.189)} &	3.114	(0.181)\\\hline
				
				RTE		& BIC	& 0.319	(0.014) &	0.406	(0.007) &	0.246	(0.011) &	0.140	(0.002) &	0.130	(0.008) &	0.171	(0.013) &	\textbf{0.081	(0.007)} \\
				& CV	& 0.230	(0.018) &	0.251	(0.019) &	0.223	(0.012) &	0.260	(0.001) &	0.219	(0.010) &	0.085	(0.031) &	\textbf{0.061	(0.007)}\\\hline
				
				uMCC	&BIC	& 0.879	(0.017) &	0.869	(0.011) &	0.933	(0.010) &	0.878	(0.011) &	\textbf{0.936	(0.008)} &	0.874	(0.011) &	0.917	(0.013)
				\\
				& CV	& 0.744	(0.019) &	0.664	(0.012) &	0.798	(0.032) &	0.892	(0.011) &	\textbf{0.954	(0.009)} &	0.804	(0.021) &	0.815	(0.007)
				
				\\\hline
				
				BA		&BIC	& 0.987	(0.004) &	0.987	(0.004) &	0.979	(0.007) &	0.792	(0.018) &	0.891	(0.015) &	\textbf{0.988	(0.004)} &	0.983	(0.006)
				
				\\
				& CV	&0.976	(0.006) &	0.945	(0.007) &	0.983	(0.005) &	0.815	(0.018) &	0.969	(0.009) &	0.984	(0.003) &	\textbf{0.986	(0.003)}
				\\\hline

				\hline
				\hline
			\end{tabular}
		}
	\end{center}
\end{table}

\newpage

\section{Asymptotic analysis}
\label{AppendixD}

In this section, we analyze the convergence rate of the proposed estimator $\widehat{\Omega}_{\text{EAGL}}$. We make the following assumptions to guarantee the existence of the true precision matrix $\Omega$:
\begin{itemize}
	\item[A1]: $\lambda_{\min}(\Omega)\ge \underline{\lambda}>0$,
	\item[A2]: $\lambda_{\max}(\Omega)\le \bar{\lambda}$,
\end{itemize}
where $\lambda_{\min}(\Omega)$ and $\lambda_{\max}(\Omega)$ are the smallest and largest eigenvalues of $\Omega$, respectively, and $\bar{\lambda}$ and $\underline{\lambda}$ are some positive values. We define the support set $S=\{(i,j):\Omega_{ij}\not=0\}$ and assume that $\text{card}(S)\le s$. 

Under the assumptions $A1$ and $A2$, if $\gamma \asymp \sqrt{\dfrac{\log p}{n}} $, then
	\begin{equation}
		\label{convergence}
		||\widehat{\Omega}_{\text{EAGL}}-\Omega||_2=O_P\left(\sqrt{\dfrac{(p+s)\log p}{n}}\right),
	\end{equation}
under the standard asymptotics (i.e., assuming that $p$ remains fixed, while $n$ converges to infinity).

\vspace{0.5cm}

The proof of (\ref{convergence}) is motivated by \cite{rothman}. First, we rewrite the optimization problem of our proposed methodology as: 
  \begin{eqnarray}
 	\label{EAGL1}
 	\centering
 	\widehat{\Omega}_{\text{EAGL}}&=&\arg\min_{\Omega}\  -\left(1+(1-\alpha)\gamma\right) \log\det(\Omega)+\text{trace}(\Omega S) + \gamma\alpha ||\Omega||_{1}.
 \end{eqnarray}
 
 Consider the following function:
\begin{eqnarray*}
	\centering
	Q(\Theta)& = & -\left(1+(1-\alpha)\gamma\right) \log\det(\Theta) +  \text{trace}(\Theta S) + \gamma\alpha||\Theta||_1- \\ 
	& - & \left( - \left(1+(1-\alpha)\gamma\right)\log\det(\Omega) + \text{trace}(\Omega S)+ \gamma\alpha||\Omega||_1 \right) \\
	& = & -\left(1+(1-\alpha)\gamma\right) \left(\log\det(\Theta)-\log\det(\Omega)\right) + \text{trace}(\Theta-\Omega)S+\gamma \alpha \left(||\Theta||_1-||\Omega||_1\right),
\end{eqnarray*}
where $\Omega$ is the true precision matrix. Note that the estimator $\widehat{\Omega}_{\text{EAGL}}$ minimizes the function $Q(\Theta)$. This means that the difference matrix $\widehat{\Delta}=\widehat{\Omega}_{\text{EAGL}} -\Omega$ minimizes the function $G(\Delta)=Q(\Omega+\Delta)$. Moreover, $G(\widehat{\Delta})\le G(0) = Q(\Omega) = 0$ and $G(\Delta)$ is a convex function.

Define the following set:
\begin{align*}
	\Phi_n(M)=\{\Delta: \Delta=\Delta^T, ||\Delta||_2=M r_n\},
\end{align*}
where $r_n=\sqrt{\dfrac{(p+s)\log p}{n}}\to 0$. If we show that $\inf\{G(\Delta): \Delta\in \Phi_n(M)\}>0$, then the minimizer $\widehat{\Delta}$ must be inside the sphere defined by $\Phi_n(M)$, i.e., $||\widehat{\Delta}||_2\le Mr_n$.
\begin{eqnarray}
	\label{functionG}
	G(\Delta) & = &	Q(\Omega + \Delta) = -(1+(1-\alpha)\gamma) \left( \log\det(\Omega + \Delta)-\log\det(\Omega) \right) \\ \nonumber
	& + & \text{trace}(\Omega + \Delta -\Omega)S + \gamma \alpha \left(||\Omega + \Delta||_1-||\Omega||_1\right) \\\nonumber
	& = & -\left(1+(1-\alpha)\gamma\right) \left(\log\det(\Omega + \Delta)-\log\det(\Omega)\right) + \text{trace} (S - \Sigma) \Delta\\ 
	& + & \text{trace} \Sigma \Delta  + \gamma \alpha \left(||\Omega + \Delta||_1-||\Omega||_1\right) \nonumber
	\end{eqnarray}
From the Taylor expansion of the function $g(t)=\log\det(\Omega + t\Delta)$, we get that:
\begin{multline*}
	\log\det(\Omega+\Delta)-\log\det(\Omega)=\text{trace}(\Omega^{-1}\Delta)-\\ \text{vec}(\Delta)^T\left[\int\limits_{0}^{1}(1-\tau)(\Omega+\tau\Delta)^{-1}\otimes(\Omega+\tau\Delta)^{-1}d\tau\right]\text{vec}(\Delta),
\end{multline*}
where $\otimes$ is the Kronecker product, and $\text{vec}(\Delta)$ is a vectorization of matrix $\Delta$. For $\Delta\in \Phi_n(M)$, we have:
\begin{multline}
\text{vec}(\Delta)^T\left[\int\limits_{0}^{1}(1-\tau)(\Omega+\tau\Delta)^{-1}\otimes(\Omega+\tau\Delta)^{-1}d\tau\right]\text{vec}(\Delta) \ge \\ \lambda_{\min}\left(\int\limits_{0}^{1}(1-\tau)(\Omega+\tau\Delta)^{-1}\otimes(\Omega+\tau\Delta)^{-1}d\tau\right)||\Delta||_2^2 \ge  \\ \int\limits_{0}^{1}(1-\tau)\lambda_{\min}^2(\Omega+\tau\Delta)^{-1}d\tau \ ||\Delta||_2^2 \ \ \ge \ \ 
	\frac{1}{2}\min_{0\le \tau\le 1}\lambda_{\min}^2(\Omega+\tau\Delta)^{-1} ||\Delta||_2^2.
\end{multline}
Furthermore:
\begin{align}
	\min_{0\le \tau\le 1}\lambda_{\min}^2(\Omega + \tau \Delta)^{-1} \ge \lambda_{\max}^{-2}(\Omega + \Delta)\ge(||\Omega||_{\text{sp}}+||\Delta||_{\text{sp}})^{-2}\ge \dfrac{1}{\bar{\lambda}+||\Delta||_{\text{sp}}^{2}} \ge \dfrac{1}{2} \bar{\lambda}^{-2},
\end{align}
since $||\Delta||_{\text{sp}}\le||\Delta||_2=o(1)$, with probability tending to 1. 

The equation $(\ref{functionG})$ can be rewritten in the following form:
\begin{multline}
	\label{functionG1}
	G(\Delta)=-\left(1+(1-\alpha)\gamma\right)\left(\text{trace} \Sigma \Delta -  \dfrac{1}{4} \bar{\lambda}^{-2}||\Delta||_2^2 \right) + \\ \text{trace}\left(\Delta(S-\Sigma)\right) + \text{trace}\left(\Sigma \Delta\right) + \gamma \alpha \left(||\Omega+\Delta||_1-||\Omega||_1\right) = \\
	\left(1+(1-\alpha)\gamma\right)\dfrac{1}{4} \bar{\lambda}^{-2}||\Delta||_2^2 -(1-\alpha)\gamma \text{trace} \Sigma \Delta + \\  \text{trace}\left(\Delta(S-\Sigma)\right)  + \gamma \alpha \left(||\Omega+\Delta||_1-||\Omega||_1\right).
\end{multline}

For an index set $S$ and a matrix $A=[a_{ij}]$, we denote $A_S=[a_{ij}I((i,j)\in S)]$, where $I(\cdot)$ is an indicator function. Recall $S=\{(i,j):\Omega_{ij}\not=0\}$, and $\bar{S}$ is its complement. It is straightforward that $||\Omega_{\bar{S}}||_1 = 0$. From the triangular inequality, we have
\begin{multline*}
	||\Omega+\Delta||_1-||\Omega||_1= ||\Omega_S+\Delta_S||_1+||\Delta_{\bar{S}}||_1  - ||\Omega_S||_1 \ge \\ ||\Omega_S||_1 - ||\Delta_S||_1+||\Delta_{\bar{S}}||_1  - ||\Omega_S||_1 = ||\Delta_{\bar{S}}||_1 - ||\Delta_S||_1
\end{multline*}
Next, we consider the term $\text{trace}\left(\Delta(S-\Sigma)\right)$. Following the results of \cite{bickellevina}, we obtain:
\begin{align*}
	\text{trace}\left(\Delta(S-\Sigma)\right) \le ||S-\Sigma||_\infty ||\Delta||_1 = O_P\left(\sqrt{\dfrac{\log p}{n}}\right)||\Delta||_1 \le C_1\sqrt{\dfrac{\log p}{n}}||\Delta||_1,
\end{align*}
with probability tending to 1, where $C_1$ is some positive value. Note that $||\Delta||_1 = ||\Delta_{\bar{S}}||_1 + ||\Delta_S||_1$.

The lower bound of the term $\text{trace} \Sigma \Delta$ is obtained using Cauchy-Schwartz inequality:
\begin{align}
	\text{trace} \Sigma \Delta \le ||\Sigma||_2 ||\Delta||_2,
\end{align}
also with probability tending to 1. Now, take $\gamma = \dfrac{C_1}{\epsilon}\sqrt{\dfrac{\log p}{n}}$, where $\epsilon$ is a small positive value. Combining the obtained lower bounds, we get from (\ref{functionG1}):
\begin{multline*}
	\label{functionG2}
	G(\Delta)\ge \left(1+(1-\alpha)\dfrac{C_1}{\epsilon}\sqrt{\dfrac{\log p}{n}} \right)\dfrac{1}{4} \bar{\lambda}^{-2}||\Delta||_2^2 -(1-\alpha)\dfrac{C_1}{\epsilon}\sqrt{\dfrac{\log p}{n}} ||\Sigma||_2 ||\Delta||_2 -\\  C_1\sqrt{\dfrac{\log p}{n}}\left(||\Delta_{\bar{S}}||_1 + ||\Delta_S||_1\right)  + \dfrac{C_1}{\epsilon}\sqrt{\dfrac{\log p}{n}} \alpha \left(||\Delta_{\bar{S}}||_1 - ||\Delta_S||_1 \right) =  \\ \left(1+(1-\alpha)\dfrac{C_1}{\epsilon}\sqrt{\dfrac{\log p}{n}} \right)\dfrac{1}{4} \bar{\lambda}^{-2}||\Delta||_2^2 -(1-\alpha)\dfrac{C_1}{\epsilon}\sqrt{\dfrac{\log p}{n}} ||\Sigma||_2 ||\Delta||_2 -\\ ||\Delta_{\bar{S}}||_1C_1\sqrt{\dfrac{\log p}{n}} \left(1-\dfrac{\alpha}{\epsilon}\right) - ||\Delta_S||_1 C_1\sqrt{\dfrac{\log p}{n}}  \left(1 + \dfrac{\alpha}{\epsilon}\right) \ge \\ \left(1+(1-\alpha)C_1\sqrt{\dfrac{\log p}{n}} \right)\dfrac{1}{4} \bar{\lambda}^{-2}||\Delta||_2^2 -(1-\alpha)C_1\sqrt{\dfrac{\log p}{n}} ||\Sigma||_2 ||\Delta||_2 - ||\Delta_S||_1 C_1\sqrt{\dfrac{\log p}{n}}  \left(1 + \dfrac{\alpha}{\epsilon}\right).
\end{multline*}
 Note that $||\Delta_S||_1\le\sqrt{s}||\Delta_S||_2\le\sqrt{s}||\Delta||_2$, which gives us
 \begin{multline*}
	G(\Delta)\ge  \left(1+(1-\alpha)\dfrac{C_1}{\epsilon}\sqrt{\dfrac{\log p}{n}} \right)\dfrac{1}{4} \bar{\lambda}^{-2}||\Delta||_2^2 -(1-\alpha)\dfrac{C_1}{\epsilon}\sqrt{\dfrac{\log p}{n}} ||\Sigma||_2 ||\Delta||_2 - \\ C_1\sqrt{\dfrac{s\log p}{n}}  \left(1 + \dfrac{\alpha}{\epsilon}\right) ||\Delta||_2 \ge \\ 
	 ||\Delta||_2^2 \biggl( \Bigl(1+(1-\alpha)\dfrac{C_1}{\epsilon}\sqrt{\dfrac{\log p}{n}} \Bigl)\dfrac{1}{4} \bar{\lambda}^{-2} -(1-\alpha)\dfrac{C_1}{\epsilon}\sqrt{\dfrac{(s+p)\log p}{n}} ||\Sigma||_2 ||\Delta||_2^{-1}  - \\   C_1\sqrt{\dfrac{(s+p)\log p}{n}}  \left(1 + \dfrac{\alpha}{\epsilon}\right) ||\Delta||_2^{-1} \biggl) = ||\Delta||_2^2 \biggl( \Bigl(1+(1-\alpha)\dfrac{C_1}{\epsilon}\sqrt{\dfrac{\log p}{n}} \Bigl)\dfrac{1}{4} \bar{\lambda}^{-2} - \\
	 (1-\alpha)\dfrac{C_1}{M\epsilon} ||\Sigma||_2 -   \dfrac{C_1}{M}\left(1 + \dfrac{\alpha}{\epsilon}\right) \biggl),
\end{multline*}
which is positive for a sufficiently large $M$. This establishes the convergence rate $(\ref{convergence})$. An open question remains regarding the degree of stringency of the positivity of $G(\Delta)$ depending on $\alpha$ and the limitations on $||\Sigma||_2$. We leave this problem to be studied in the future.

\end{document}